\def\NAME#1#2{\caption{#2} \label{#1}}

\def\SEC#1#2{\section{#1} \label{#2}}
\def\SUB#1#2{\subsection{#1} \label{#2}}

\def\FIGT#1{\begin{figure}[!t] \begin{center} #1 \end{center} \end{figure}}

\def\FIGWT#1{\begin{figure*}[!t] \begin{center} #1 \end{center} \end{figure*}}

\def\TABLET#1{\begin{table}[!t] \begin{center} #1 \end{center} \end{table}}
\def\TABLEB#1{\begin{table}[!b] \begin{center} #1 \end{center} \end{table}}

\def\EPSF#1#2#3#4{\includegraphics[#3= #4 cm,clip]{#1.eps} \NAME{#1}{#2}}

\def\EPS08#1#2#3#4{\includegraphics[#3= #4 cm,clip]{../eps08/#1.eps} \NAME{#1}{#2}}

\def\EQ#1{\begin{equation} #1 \end{equation}}
\def\EQL#1#2{\begin{equation} \label{#1} #2 \end{equation}}
\def\EQN#1{\begin{eqnarray} #1 \end{eqnarray}}
\def\SP#1{\begin{split} #1 \end{split}}
\def\AL#1{ \begin{array}{l} #1 \end{array} }

\def\LRA#1{ \langle #1 \rangle}
\def\LP#1{\left \{ \AL{#1} \right.}
\def\LPS#1{\left \{ \AL{ \SP{#1} } \right.}

\newfont{\bg}{cmr10 scaled\magstep5}
\newfont{\bbg}{cmsy10 scaled\magstep5}
\newcommand{\bigzerol}{\smash{\hbox{\bg 0}}}
\newcommand{\bigzerou}{\smash{\lower2.7ex\hbox{\bg 0}}}

\documentclass[twocolumn,showpacs,preprintnumbers,amsmath,amssymb]{revtex4}

\usepackage{graphicx}
\usepackage{dcolumn}
\usepackage{bm}

\usepackage[breaklinks]{hyperref}

\begin{document}


\title{Introduction of Frictional and Turning Function for \\ Pedestrian Outflow with an Obstacle}

\author{Daichi Yanagisawa$^{1,2}$}
\email{tt087068@mail.ecc.u-tokyo.ac.jp}

\author{Ayako Kimura$^3$}
\author{Akiyasu Tomoeda$^{4,1}$}
\author{Ryosuke~Nishi$^1$}
\author{Yushi Suma$^1$}
\author{Kazumichi Ohtsuka$^1$}

\author{Katsuhiro Nishinari$^{1,5}$}
 
\affiliation{%
$^1$Department of Aeronautics and Astronautics, School of Engineering, The University of Tokyo, Hongo, Bunkyo-ku, Tokyo, 113-8656, Japan. \\
$^2$ Research Fellow of the Japan Society for the Promotion of Science, Kojimachi, Chiyoda-ku, Tokyo 102-8471, Japan \\
$^3$Mitsubishi Research Institute, Inc., \\
Otemachi, Chiyoudaku, Tokyo, 100-8141, Japan \\
$^4$Meiji Institute for Advanced Study of Mathematical Sciences, Meiji University, Higashimita, Tama-ku, Kawasaki-city, Kanagawa, 214-8571, Japan \\
$^5$ PRESTO, Japan Science and Technology Agency, Sanbanmachi, Chiyoda-ku, Tokyo 102-0075, Japan \\
}%


\date{\today}

\begin{abstract}
In this paper, two important factors which affect the pedestrian outflow at a bottleneck significantly are studied in detail to analyze the effect of an obstacle set up in front of an exit.
One is a conflict at an exit when pedestrians evacuate from a room.
We use floor field model for simulating such behavior, which is a well-studied pedestrian model using cellular automata.
The conflicts have been taken into account by the friction parameter.
However, the friction parameter so far is a constant and does not depend on the number of the pedestrians conflicting at the same time.
Thus, we have improved the friction parameter by the \textit{frictional function}, which is a function of the number of the pedestrians involved in the conflict.
Second, we have newly introduced the cost of \textit{turning} of pedestrians at the exit.
Since pedestrians have inertia, their walking speeds decrease when they turn, and the pedestrian outflow decreases.

The validity of the extended model, which includes the frictional function and the turning function, is verified by both a mean field theory and experiments.
In our experiments, the pedestrian flow increases when we put an obstacle in front of an exit.
The analytical results clearly explains the mechanism of the effect of the obstacle, i.e., the obstacle blocks pedestrians moving to the exit and decreases the average number of pedestrians involved in the conflict.
We have also found that an obstacle works more effectively when we shift it from the center since pedestrians go through the exit with less turning. 
\end{abstract}


\maketitle


\SEC{Introduction}{INTRO}
Pedestrian Dynamics has been studied vigorously in physics field over last decades \cite{socialreview,nagatani}.
Many microscopic models are developed such as the floor field (FF) model \cite{p-matrix, dff, competition, friction,  inertia-wall, force, RCA, dy1pre, exit-dens}, the social force model \cite{socialreview,social, socialnature}, and the lattice-gas model \cite{nagatani,multigrid, largeobject}, to simulate a crowd of pedestrians movement realistically.
In addition to the simulations, there are also many pedestrian dynamics experiments \cite{tgf03-Hoogendoorn, ex_fundamental1, ex_bottleneck1, ex_bottleneck2} to study collective behaviors of pedestrians, while individual characteristics have been experimentally studied in psychology and physiology \cite{ex_turn1}.

\FIGWT{\EPSF{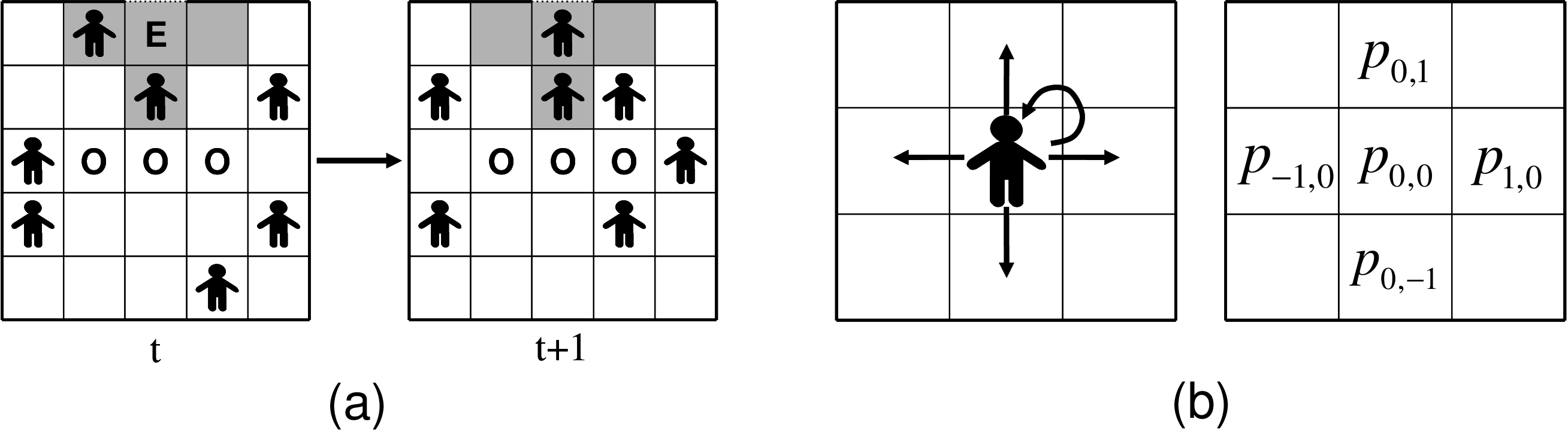}{
(a) A schematic view of an evacuation simulation by the FF model. 
The person-shaped silhouettes represent pedestrians; the letters \textbf{E} and \textbf{O} represent the exit cell and obstacle cells, respectively.
Pedestrians proceed to the exit by one cell at most by one time step.
(b) Target cells for a pedestrian at the next time step. The motion is restricted to the Neumann neighborhood in this model.
}{height}{4.7}}

The behavior of pedestrians at an exit is focused on by many researchers in physics since it greatly affects the total evacuation time in an emergency situation \cite{dy1pre, ex_bottleneck1, ex_bottleneck2}.
Researchers of multi-robot systems have also studied such evacuation \cite{robot1}. 
In the Ref. \cite{friction}, it is indicated that we can evacuate faster when an obstacle is put in front of an exit.
In this paper, we have extended FF model for an analysis on the pedestrian outflow with an obstacle by introducing two important factors of pedestrian dynamics at an exit . 
One is a \textit{conflict} and the other is a \textit{turning}.
A conflict occurs when more than one pedestrian move to the same place at the same time.
When many conflicts occur, the pedestrian outflow, which is the number of pedestrians going through an exit with unit width per unit time, decreases, and the total evacuation time increases.
Moreover, pedestrians must change their proceeding direction to the exit, when they go through it.
Since pedestrians have inertia, although it is much smaller than that of vehicles, their walking speeds decreases when they turn.
This decrease in walking speed results in the decrease of the pedestrian outflow.

In the FF model, which is a pedestrian dynamics model using cellular automata, conflicts are take into account by \textit{friction parameter} \cite{competition}, which describes clogging and sticking effects between pedestrians. 
Since the friction parameter is a constant parameter, the strength of clogging and sticking does not depend on the number of pedestrians involved in the conflict.
In reality, however, it is more difficult to avoid a conflict when three pedestrians move to the same place at the same time than two pedestrians move to the same place at the same time.
The inertia of pedestrians is considered in Ref. \cite{inertia-wall,turn1,Tobias_Kretz}, however, this inertia effect is the suppression of quick changes of the direction of the motion, i.e., pedestrians try to keep walking in the same direction.
The inertia effect, which should be introduced at an exit, is not the suppression of quick changes, but the decrease in walking speed, which has not been considered in the previous models.

Therefore, we newly introduced the \textit{frictional function} and the \textit{turning function}.
Frictional function is a function of the number of pedestrians involved in the conflict, which reproduces the difference of the strength of clogging against it.
Turning Function is a function of the angle of deviation from the former direction of motion, which estimates the decrease in walking speed quantitatively by turning.
These functions make it possible to describe the pedestrian behavior around an exit more precisely and realistically. 
The mechanism of the effect of the obstacle is also explained successfully by the frictional and turning functions.

We have also performed the experiments of evacuations and verified that the outflow obtained by the extended model agrees with the experimental data well.
The outflow increases when an obstacle is put at the proper position in front of the exit as we have expected from our theoretical calculation.

We would like to emphasize that our study is based on firm theoretical analysis.
Vehicle traffic has been theoretically studied by extending asymmetric simple exclusion process \cite{C&S&S,lagrange-nishinari,DW-H1,DW-H2}, however, pedestrians dynamics has been hardly done because of the complexity of the rules of motion and the two-dimensionality.
Therefore, theoretical analysis on pedestrian dynamics is also a new challenge.

This paper is organized as follows.
In Sec. \ref{FFM}, we briefly review the FF model.
The two new functions, i.e., the frictional function and the turning function, are introduced in Sec. \ref{FFUNC} and Sec. \ref{TFUNC}, respectively.
In Sec. \ref{NNC}, several kinds of lattice for FF model is considered. 
We obtain the analytical expression of the average pedestrian outflow by using the cluster approximation in Sec. \ref{AOCA}.
In Sec. \ref{EXP1} and Sec. \ref{EXP2}, we show the result of the evacuation experiments in lines and the evacuation experiment with an obstacle, respectively, and investigate the mechanism of increasing outflow by obstacle.
Section \ref{CONC} is devoted to summary and discussion.


\SEC{Floor Field Model}{FFM}

\subsection{Floor Field}
We consider a situation that every pedestrian in a room moves to the same exit.
The room is divided into cells as given in Fig. \ref{pre1_intro} (a).
Man shaped silhouettes represent pedestrians, an alphabet \textbf{E} and alphabets \textbf{O} represent the exit cell and obstacle cells, respectively.
Each cell contains only a single pedestrian at most.
Every time step pedestrians choose which cell to move from 5 cells for simplicity: a cell which the pedestrian stands now [$(i,j)=(0,0)$] and the Neumann neighboring cells [$(i,j)=(0,1)$, $(0,-1)$, $(1,0)$, $(-1,0)$] (Fig. \ref{pre1_intro} (b)).
Two kinds of FFs determine the probability of which direction to move, i.e., Static FF $S_{ij}$, which is the shortest distance to the exit cell (Fig. \ref{pre2_sff}), and Dynamic FF $D_{ij}$, which is a number of footprints left by the pedestrians \cite{dff}.
Dynamic FF plays role in mimicking the long-raged interaction among pedestrians to short-ranged one.
The transition probability $p_{ij}$ for a move to a neighboring cell $(i,j)$ is determined by the following expression,
\begin{equation}
p_{ij}=N\xi _{ij}\exp (-k_sS_{ij} +k_{d}D_{ij}). \label{pp}
\end{equation}
Here the values of the FFs $S_{ij}$ and $D_{ij}$ at each cell $(i,j)$ are weighted by two sensitivity parameters $k_s$ and $k_d$ with the normalization $N$.
There is a minus sign before $k_s$ since pedestrian move to a cell which Static FF decreases.
$\xi_{ij}$ returns 0 for an obstacle or a wall cell and returns 1 for other kinds of cells.
Note that in our paper a cell occupied by a pedestrian is not regarded as an obstacle cell, thus it affects the normalization $N$.
In the following, we ignore the effect of Dynamic FF, i.e., $k_d=0$, since we study the pedestrian behavior at an exit, where there is only short-ranged interaction.

\FIGT{\EPSF{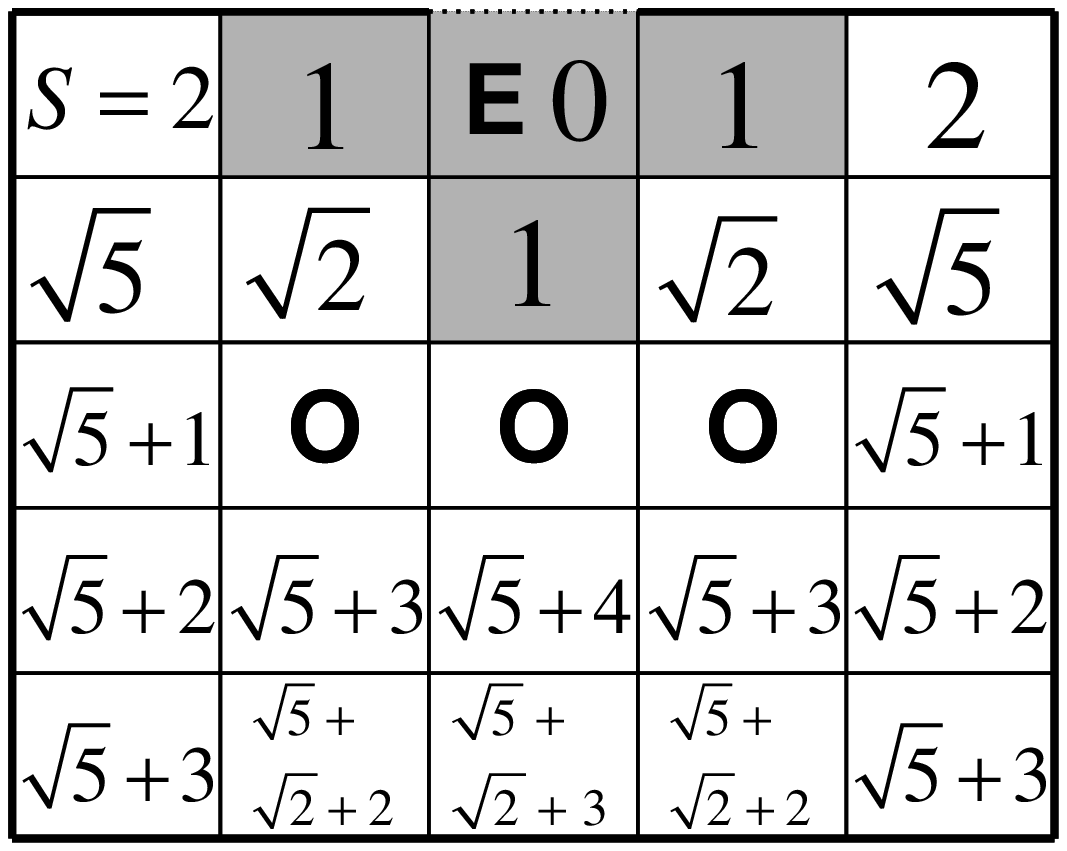}{
Static floor field constructed by the exit \textbf{E}. The numbers in each cell represent the Euclidean distances from the exit cell, which are calculated by Dijkstra method \cite{dijkstra}.
}{height}{5}}

\subsection{Bottleneck Parameter} \label{bottleneck}
We use bottleneck parameter $\beta\in[0,1]$, which was introduced in Ref. \cite{dy1pre}, when we consider the congested evacuation situation at an exit.  
Then, the transition probability of pedestrians who occupy one of the neighboring cells of the exit cell is transformed from $p_{ij}$ to $(p_{ij})_{\beta}$ as follows:
\begin{eqnarray}\label{pbeta}
\left \{
\begin{array}{l}
(p_{ij})_{\beta}=\beta p_{ij} \hspace{1cm} [(i, j)\not = (0, 0)],\\
(p_{0,0})_{\beta} = \beta p_{0,0} + (1 - \sum_{i,j} \beta p_{ij}) = \beta p_{0,0} + (1-\beta).
\end{array}
\right.
\end{eqnarray}
$\beta$ controls the velocity of the pedestrians who are at the neighboring cells of the exit.
When $k_s$ is large, the transition probabilities of pedestrians at neighboring cells of the exit are approximated by using $\beta$ \cite{dy1pre}.
This simplification enables us to analyze the pedestrian behavior theoretically.

\subsection{Friction Parameter} \label{collision}
Due to the use of parallel dynamics it happens that two or more pedestrians choose the same target cell in the update procedure.
Such situations are called conflicts in this paper. 
To describe the dynamics of a conflict in a quantitative way, \textit{friction parameter} $\mu$ was introduced in Refs. \cite{competition}.
This parameter describes clogging and sticking effects between the pedestrians.
When we denote the number of pedestrians moving to the same cell at the same time as $k\in\textbf{N}$, the friction parameter is not used in $k=1$, but applied in $k\geq2$ as follows:
\EQN{ \label{phim}
\phi_\mu(k) = 
\LPS{
0 \ \ \ \ \ \ \ &(k=1), \\
\mu \ \ \ \ \ \ \ &(k\geq 2),
}}
where $\phi_\mu$ is the probability that all the pedestrians involved in the conflict remain at their cell.
$\phi_\mu(1)=0$ since there is no conflict in $k=1$.
In $k\geq2$, the movement of \textit{all} involved pedestrians is denied with probability $\mu$, therefore, the conflict is solved with probability $1-\mu$, and one of the pedestrians is allowed to move to the desired cell (Fig. \ref{pre3_col}).
The pedestrian which actually moves is then chosen randomly with equal probability.
In a situation with large $\mu$, pedestrians are competitive and do not give way to others.
Thus they hardly move due to the conflict between them.
Contrary in a situation with small $\mu$ they give way and cooperate each other.

\FIGT{\EPSF{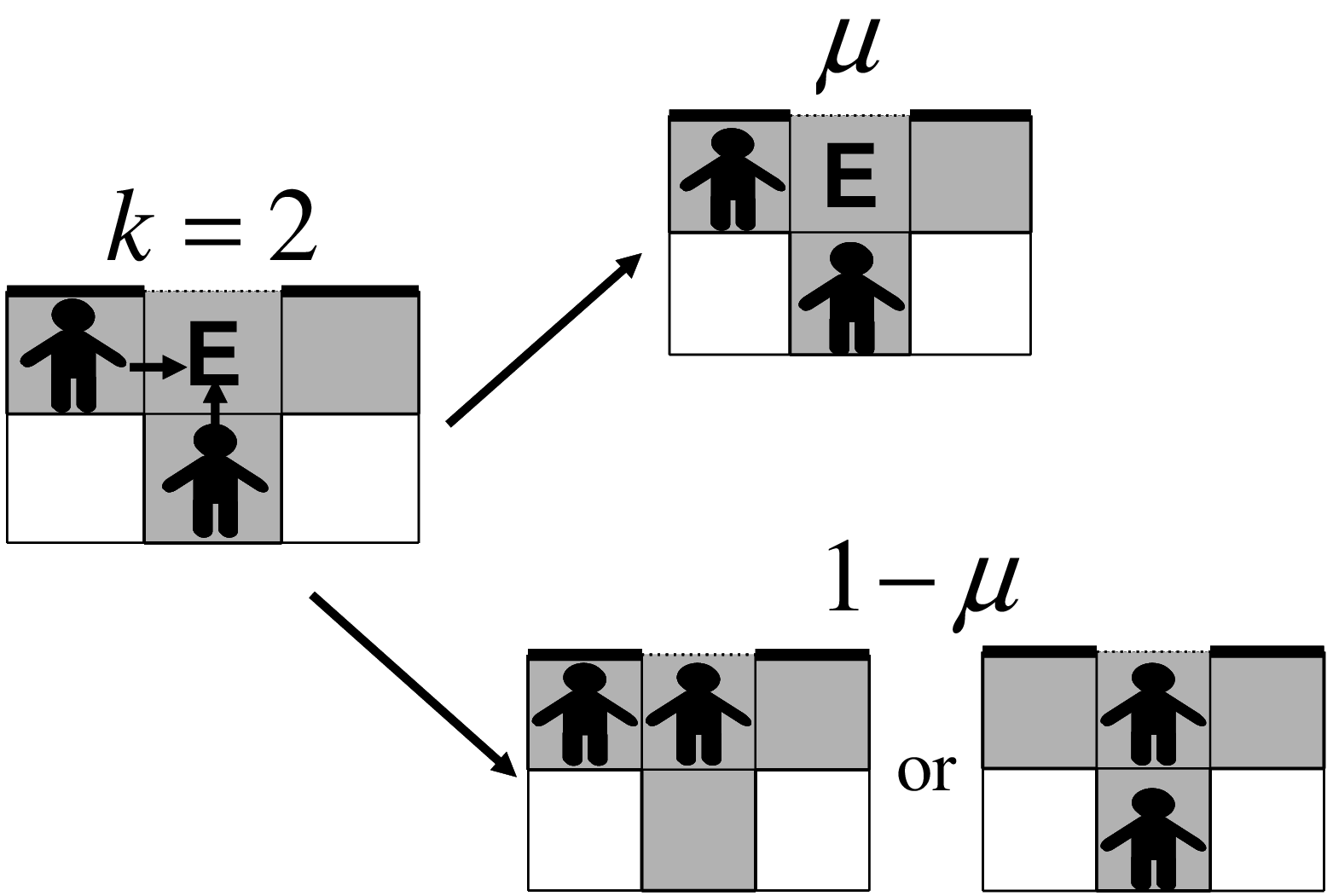}{
The way of solving conflicts. In a conflict situation $(k\geq2)$, movement of all involved pedestrians is denied with probability $\mu$. One of them is randomly allowed to move to the desired cell with probability $1-\mu$.
}{height}{5}}

\subsection{Update rules} \label{update}

The FF model which ignores the effect of the Dynamic FF consists of the following two steps per unit time step.
\begin{enumerate}
\item Calculate each pedestrian's transition probability by (\ref{pp}) and (\ref{pbeta}).
Transition probabilities of pedestrians who stand on the exit cells, i.e., the probabilities of getting out from the room, is $\alpha\in[0,1]$.
\item Move pedestrians based on the calculated transition probability with parallel update.
If there are cells which are possibly occupied by more than one pedestrian, solve conflicts by the means of Sec. \ref{collision}.
\end{enumerate}

\SEC{Frictional Function}{FFUNC}

\FIGT{\EPSF{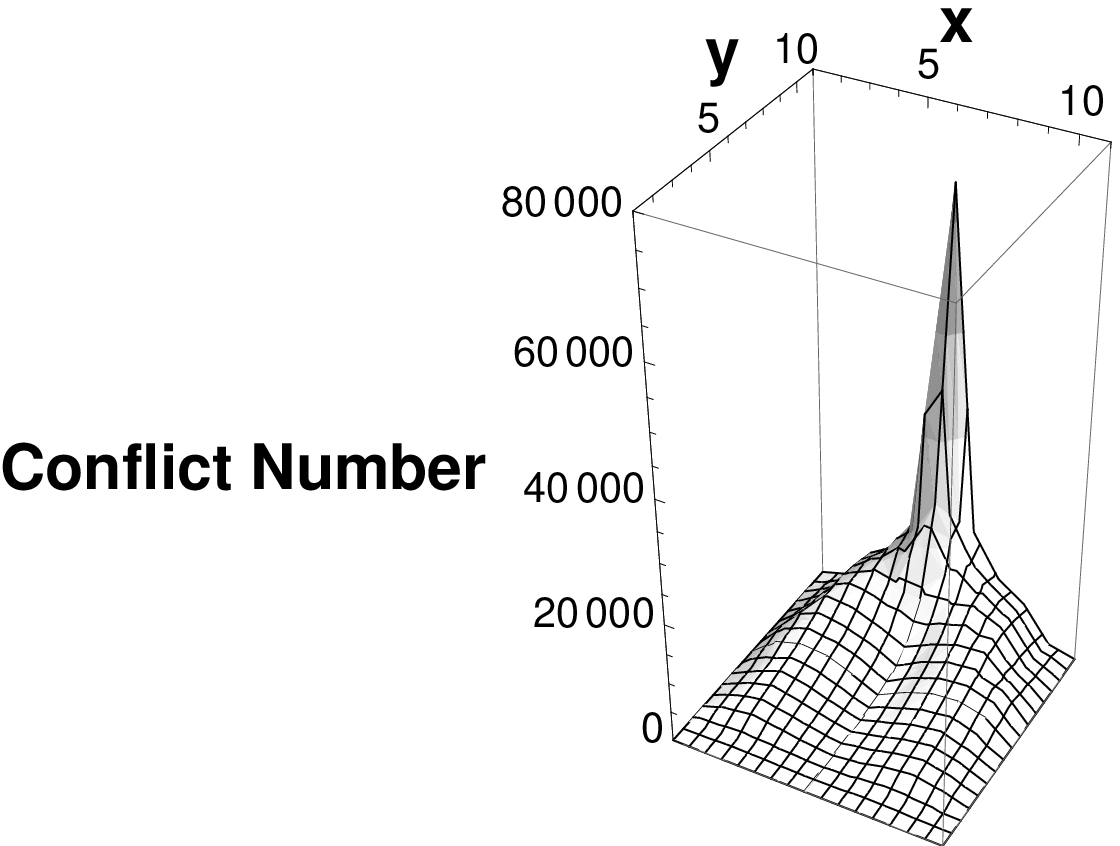}{
The number of conflicts in a competitive situation, i.e., $k_s=20$, $\beta=1$, and $\mu=0.6$. We simulated in the 11$\times$11 cells' room with the 1 cell's exit for 101,000 time steps and accumulated the value of 1,001 to 101,000 time steps. The exit cell is at $(6,10)$ and Neumann Neighborhood is used.
}{height}{5.8}}

Figure \ref{pre4_colnum} shows the number of conflicts, which is a result of a evacuation simulation using Neumann Neighborhood in 100,000 time steps.
In this simulation pedestrians enter into the room with probability 1 from the boundary cells of three sides of the square room that do not include the exit cell, therefore the room is always filled with pedestrians \cite{dy1pre}.
At the exit cell, there are 69,385 conflicts in 100,000 time steps, i.e., the probability of a conflict at there is approximately, 70\%.
Thus, the conflict is the important factor when we consider the evacuation from the narrow exit, and the friction parameter has played a role to represent it.
However, it is constant in $k\geq2$ and does not take account of the difference of the number of the pedestrians involved in the conflict.
In reality, an impact of a conflict is stronger when three pedestrians conflict each other than when two pedestrians conflict each other.
In the simulation, 85\% of the conflicts occur in the room are the conflicts between two pedestrians, and 15\% of them are the conflicts among three pedestrians.
When we consider the conflicts at the exit cell, 34\% are between two pedestrians and 66\% are among three pedestrians.
Thus, it is impossible to approximate as ``Number of the conflicts between two pedestrians" $\ll$ ``Number of the conflicts between three pedestrians" or the opposite way, and we cannot consider the only one kind of conflict.
 
Therefore, we newly introduce the \textit{frictional function} $\phi_\zeta\in[0,1]$, which is a function of the number of pedestrians moving to the same cell at the same time $k$ in general as follows:
\EQ{
\phi_\zeta(k) = 1-(1-\zeta )^k - k\zeta (1-\zeta )^{k-1} \hspace{0.5cm} (k \geq 1).  \label{phiz}
}
$\phi_\zeta$ represents the probability of the unsolved conflict and is defined by considering the psychological effect of pedestrians.
$\zeta\in[0,1]$ is an \textit{aggressive parameter}, which is a probability of not giving way to others when more than one pedestrian move to the same cell at the same time.
The term $(1-\zeta )^k$ in the expression (\ref{phiz}) is the probability that all pedestrians involved in the conflict try to give way to others.
The term $k\zeta (1-\zeta )^{k-1}$ is the probability that only one pedestrian does not give way to others while the others do.
By subtracting the two terms, which are the probabilities of resolving the conflict, from 1, we obtain the frictional function $\phi_\zeta$.
Therefore $\phi_\zeta$ includes the effect of motivations of pedestrians.

We compare the characteristics of $\phi_\zeta$, i.e., \textit{frictional function} and $\phi_\mu$, which we call \textit{frictional parameter} in the following.
$\phi_\zeta$ satisfies the two conditions:
\EQN{ 
\LP{
\phi_\zeta(1) = 0, \label{1ha0} \\
\phi_\zeta(\infty) = 1 \hspace{0.5cm} (\zeta \not= 0). \label{mugenha1}
}}
The former equation means that there is no conflict when only one pedestrian move to the target cell.
The latter equation means that no one can move to the same cell when greatly many pedestrians move to there at the same time since pedestrians have some finite volume.
$\phi_\mu$ satisfies the former equation, however if $\mu \not= 1$, it does not satisfy the latter one, which is a natural condition near the exit.

Figure \ref{pre5_mu012} is the plots of $\phi_\mu$ and $\phi_\zeta$ as a function of $k$.
$\phi_\mu$ is constant in $k\geq 2$, while $\phi_\zeta$ gradually increases as $k$ increases and reflects the difference of the strength of clogging against $k$.
We compare the validity of $\phi_\mu$ and $\phi_\zeta$ by applying them to the experimental results in the Sec. \ref{EXP1} and \ref{EXP2}.

\FIGT{\EPSF{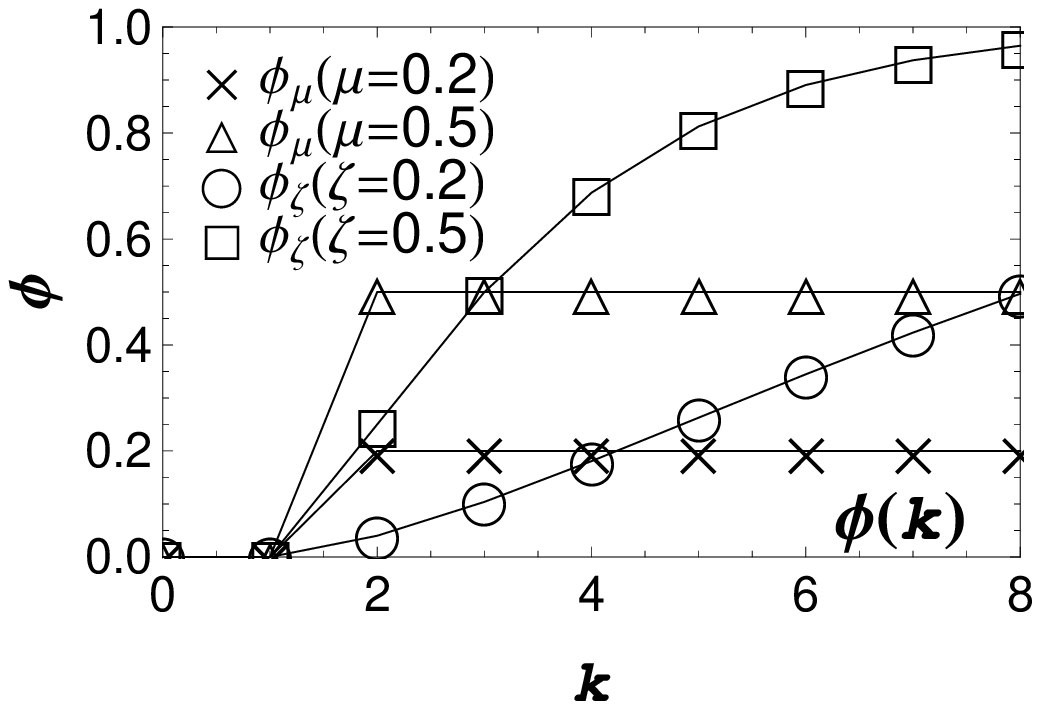}{
Values of the frictional parameter and the frictional function against $k$, which is a number of pedestrians who move to the same cell at the same.
We see that $\phi_\mu$ is constant in $k\geq 2$, while $\phi_\zeta$ increases gradually and saturate in 1.
$\phi_\mu$ and $\phi_\zeta$ become large according to the increase in $\mu$ and $\zeta$, respectively.
}{height}{5.8}}

The advantage of the frictional function is that its value depends on the number of pedestrians involved in a conflict, but does not depend on the type of the lattice, which we adopt.
Thus, we can use them in every discrete model of self-driven particles with parallel update.
We would also like to emphasize that the frictional function is not only effective for considering a narrow exit, i.e., an exit with one cell, but also a wide exit, i.e., an exit with plural cells, since the friction parameter is also applied to calculate an outflow through a wide exit in Ref. \cite{dy1pre}.

\SEC{Turning Function}{TFUNC}

\FIGT{\EPSF{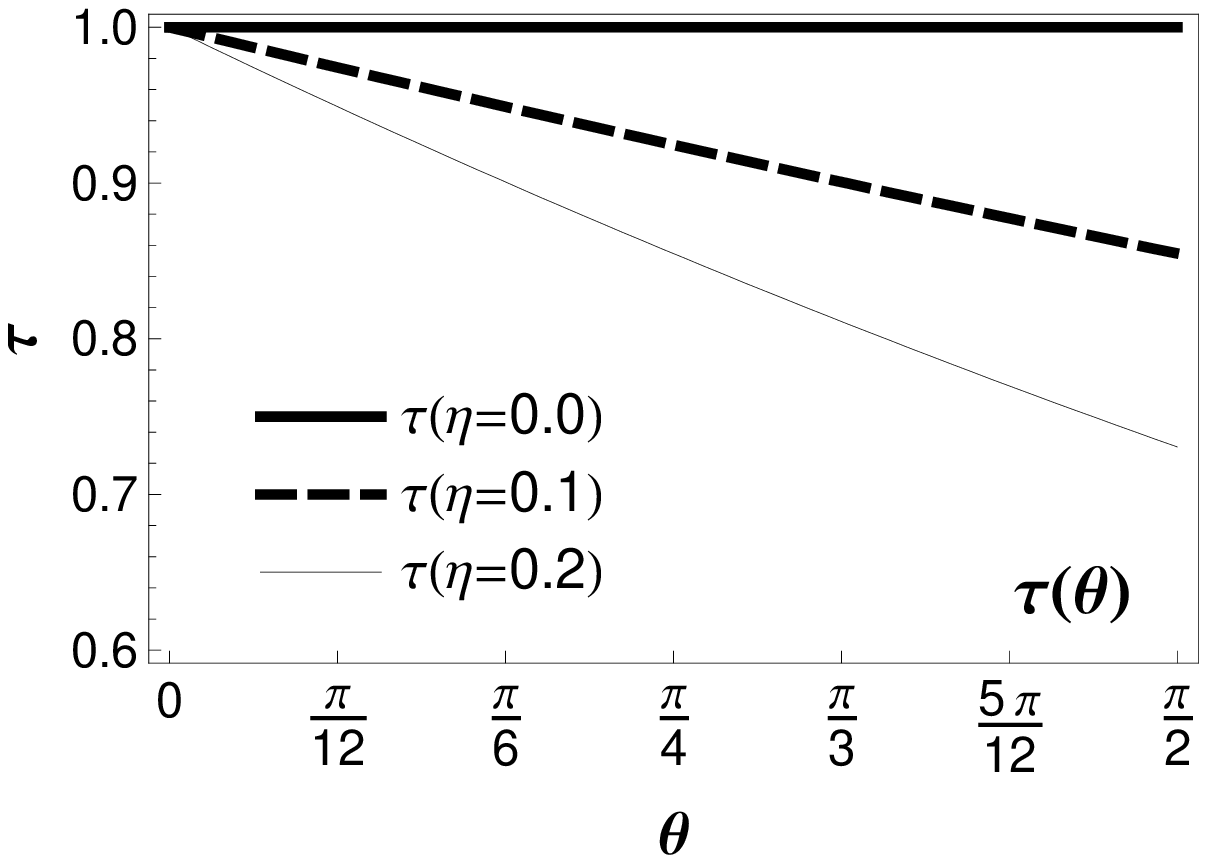}{
Values of the turning function $\tau$ against $\theta$, which is an angle of deviation from the former direction of motion.
It decreases as $\theta$ increases and its slope become sharp when $\eta$ increases. 
}{height}{5.5}}

Since pedestrians have inertia, their walking speeds decrease when they turn and change their headings.
While the inertia effect of vehicles is considered in Ref. \cite{slowstart1,slowstart2,lagrange-nishinari} as a \textit{slow-start rule}, that of pedestrian dynamic is introduced to the FF model in Ref. \cite{inertia-wall,turn1,Tobias_Kretz}.
However, these inertia effects are not appropriate for estimating the decrease in walking speed.

The inertia parameter introduced in Ref. \cite{inertia-wall} is too simple for analyzing the pedestrian behavior at the exit in detail.
When the strength of inertia increases in the model, pedestrians try to move the same direction as they moved in the previous step.
Therefore, the pedestrian move to the exit from the lateral direction
often cannot stop at the exit and pass in front of it.
Moreover, the strength of inertia does not depend on the angle between the direction of the movement in previous step and that in present step.
For instance, the probability of turning $90^\circ$ and that of turning $180^\circ$ are same.
In reality, however, it becomes more difficult to turn as the angle of deviation from the former direction becomes large.
It is also verified by the experiments in Ref. \cite{ex_turn1}.

In the F. A. S. T. model \cite{turn1,Tobias_Kretz}, an inertia effect is studied in detail.
The strength of the inertia effect is a function of the angle of deviation from the former direction and calculated by considering centrifugal forces.
However, this inertia effect is not used to decide the walking speed but only the moving direction. 

Thus, we introduce the turning function $\tau$, which represents the effect of decrease in walking speed by turning as follows:
\EQL{tau}{
\tau(\theta) = \exp \left( -\eta |\theta| \right) ,
}
where $\theta\in[\pi,-\pi]$ is the angle of deviation from the former direction of motion, and $\eta\geq 0$ are the inertia coefficient in turning, which represent the strength of the inertia.
We decide to use an exponential function for the turning function since it is simple and also matches to the transition probability (\ref{pp}) in the FF model.
An exponential function is also easy to analyze theoretically.
We see that $\tau(0)=1$ for arbitrary $\eta$ from (\ref{tau}) since there is no decrease in walking speed when pedestrians do not turn.
Figure \ref{pre6_tau123} shows that the values of the turning function, which decrease almost linearly according to the increase of the angle of deviation $\theta$ since $\eta$ is not large.
We verify that $\eta$ for real pedestrians is as small as $0.1$ in Sec. \ref{EXP1} and \ref{EXP2}.
The turning function enables us to calculate more realistic outflow. 

\FIGT{\EPSF{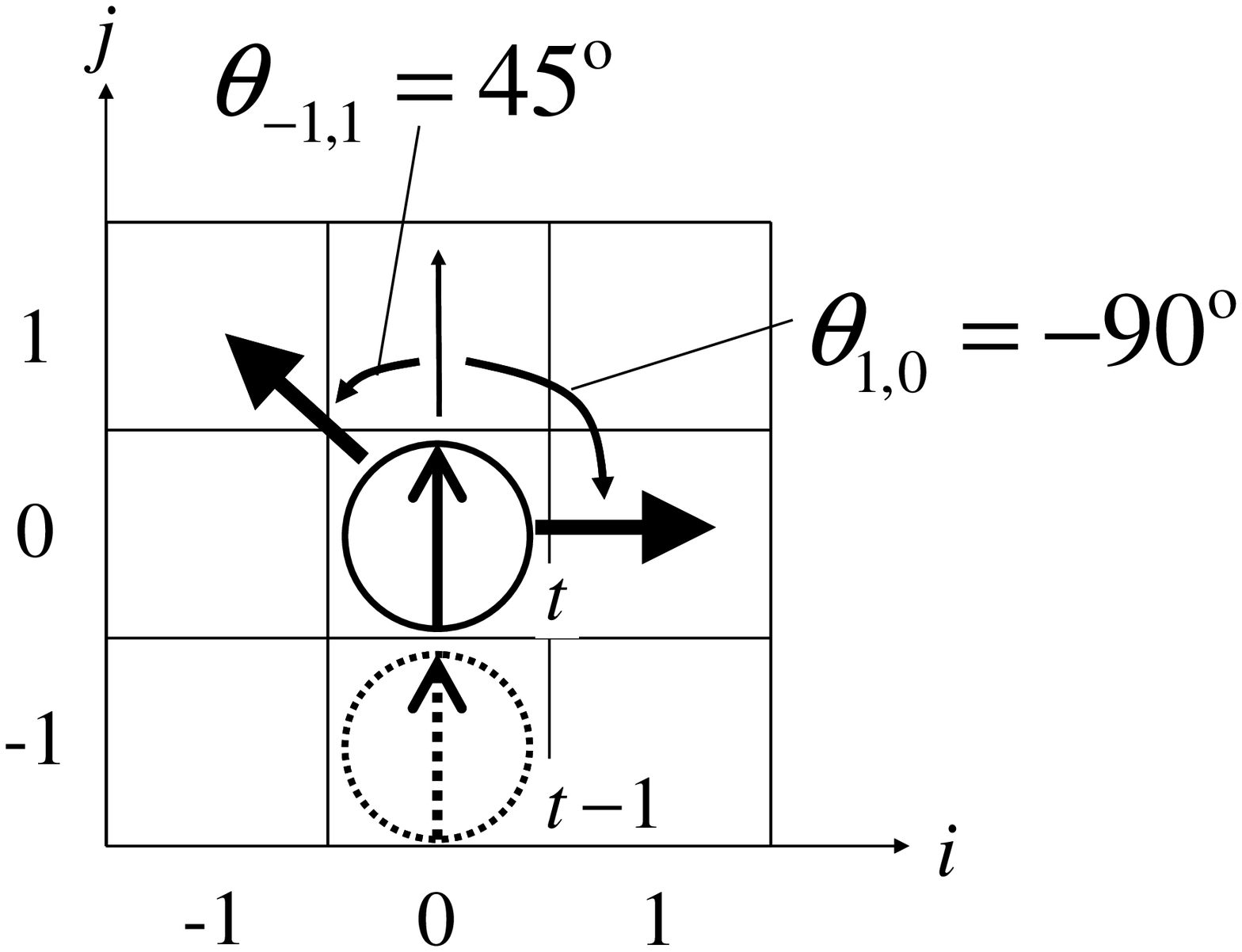}{
A schematic view of the turning pedestrian. The pedestrian, who is depicted with circled-arrow in the figure, moved to the cell $(0,0)$ from $(0,-1)$ in the time step $t-1$.
If he/she try to move to the cell $(1,0)$, he/she has to turn $\theta_{1,0}(=-90^\circ)$ to change his/her heading, and if he/she try to move to the cell $(-1,1)$, he/she has to turn $\theta_{-1,1}(=45^\circ)$ to change his/her heading.
$\theta_{ij}$ changes its value according to the former direction of the movement.
}{height}{5.5}}

$(p_{ij})_{\tau}$, which is the transition probability including the effect of turning by $\tau$, is described as,
\EQ{
\LPS{
(p_{ij})_{\tau} &= \tau(\theta_{ij}) p_{ij} \hspace{1cm} [(i, j)\not = (0, 0)],\\ \label{pturn}
(p_{0,0})_{\tau} &= \tau (0) p_{0,0} + \left( 1 - \sum_{i,j} \tau(\theta_{ij}) p_{ij} \right), \\
}
}
where $p_{ij}$ is given by (\ref{pp}), and $\theta_{ij}$ is the angle between the former direction of the motion and the direction of the cell $(i,j)$ as in Fig. \ref{pre7_turn2}.

\FIGWT{\EPSF{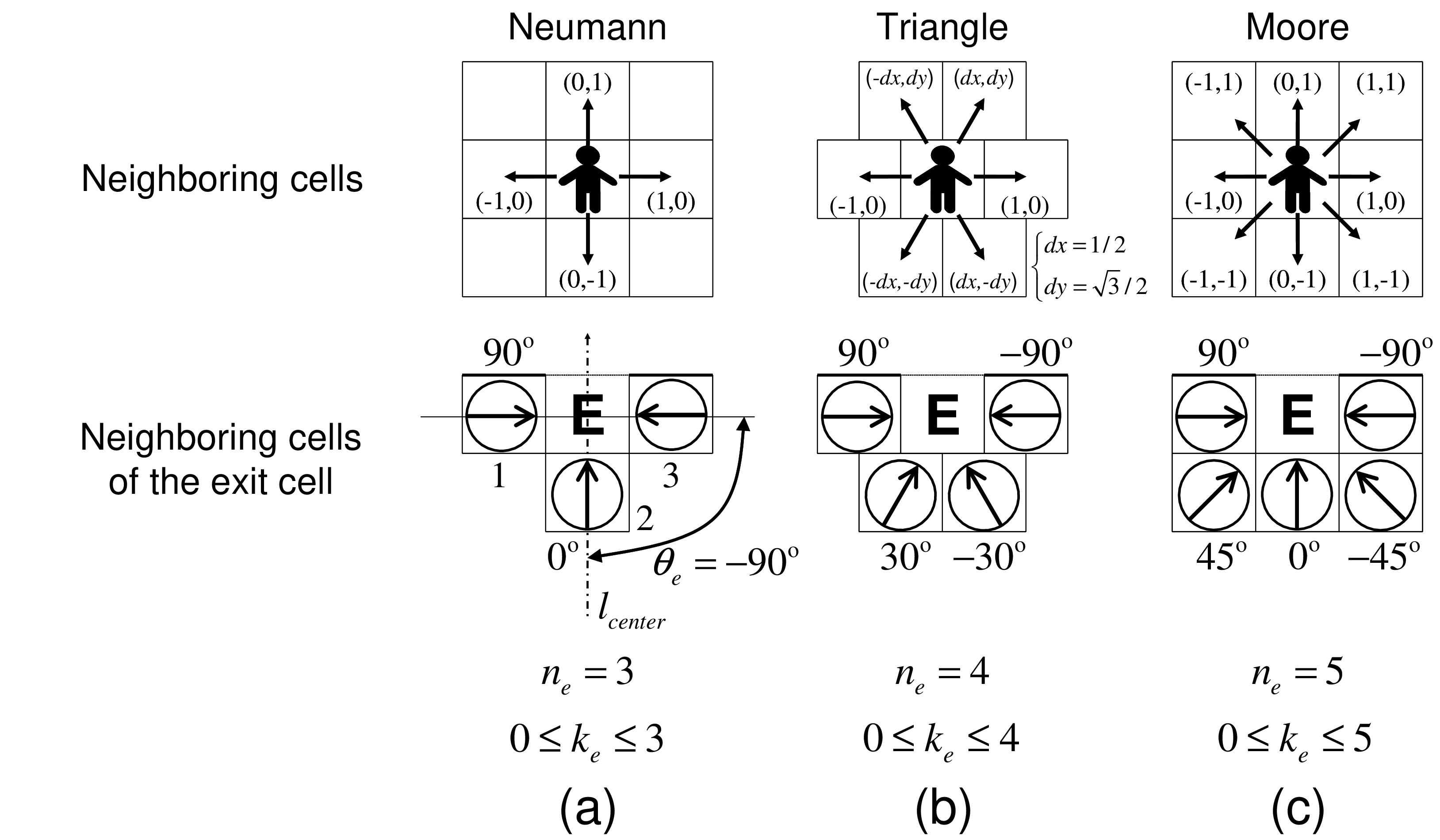}{
Schematic views of the pedestrian cluster at the exit cell for different neighborhood.
(a), (b), and (c) are the case of Neumann neighborhood, Triangle-lattice neighborhood, and Moore neighborhood, respectively.
Circled-arrows represent pedestrians and their facing direction.
In the first row, target cells for a pedestrian are depicted.
In the case of Neumann neighborhood, pedestrians can move only vertical and horizontal directions, however, they can move to diagonal ways in the case of Moore neighborhood.
Pedestrians can move to six directions in the case of Triangle-lattice neighborhood; however they cannot go straight in the vertical way. 
In the second row, the schematic views of pedestrian cluster at the exit cell and its neighboring cells are depicted.
The incident angle $\theta_e$ is defined as an angle between the line $l_{center}$ and direction of the pedestrian movement as in the figure (a).
For instance, pedestrian-3 tries to move to the exit cell from the right neighboring cell of the exit cell in the figure (a).
Thus his/her angle of incidence $\theta_{e}$ equals to $-90^\circ$.
$n_e$, i.e., the numbers of the neighboring cells of the exit cell, and ranges of $k_e$, i.e., the number of pedestrians moving to the exit at the same time are also described in the figure.
The minimum and the maximum value of $k_e$ are $0$ and $n_e$, respectively. 
}{height}{10.5}}

\SEC{Number of the Neighboring Cells and the Incident Angle}{NNC}

In this section, we consider the number of the neighboring cells of the exit cell and the incident angle to the exit.
We denote them as $n_e$ and $\theta_e$ in the following.
Since FF model adopts Neumann Neighborhood, there are three neighboring cells of the one-cell exit, i.e., $n_e=3$ (Fig. \ref{pre8_nn-ia} (a)), thus, $k_e$, which is the number of pedestrians moving to the exit at the same time, is between $0$ and $n_e(=3)$ and, there is possibility of occurring a conflict between two pedestrians and three pedestrians at the exit.
We suppose that all pedestrians at the neighboring cells face to the exit cell in the evacuation.
Then, the incident angle $\theta_e$ is defined as an angle between the vertical line perpendicular to the wall ($l_{center}$ in Fig. \ref{pre8_nn-ia} (a)) and the direction of the pedestrian movement.
Therefore, the incident angle of the pedestrian at the cell 2 in Fig. \ref{pre8_nn-ia} (a) is $0^\circ$, and those of pedestrians at the cell 1 and 3 are $90^\circ$ and $-90^\circ$, respectively.
They have to turn $\theta_e$ at the exit cell to change their orientation and go through the exit (Fig. \ref{pre9_turnseq}).

\FIGWT{\EPSF{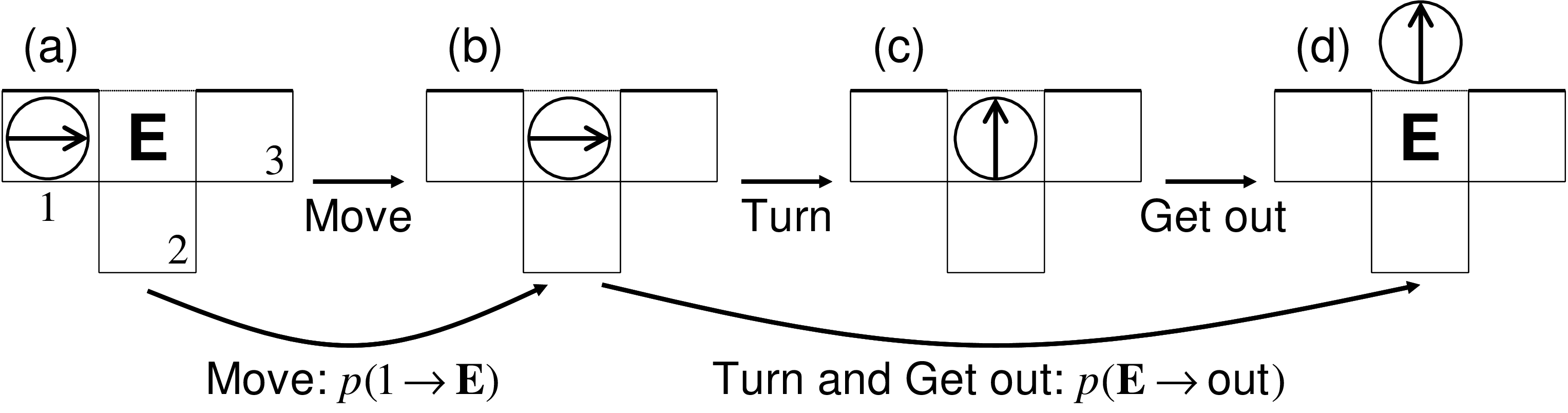}{
Schematic view of the pedestrian-1 going through the exit.
(a)$\rightarrow$(b): Move to the exit cell.
(b)$\rightarrow$(c): Turn $90^\circ$ at the exit cell.
(c)$\rightarrow$(d): Get out from the room.
The transition probability from (a) to (b) is represented by $p(1 \rightarrow \textbf{E})$ (\ref{pmEpmm}), and that from (b) to (d) is described by $p(\textbf{E} \rightarrow \rm{out})$ (\ref{alpha}).
Hence, $p(\textbf{E} \rightarrow \rm{out})$ includes both turning and getting out.
}{height}{4.5}}

Here, we consider two other neighborhoods in the FF model.
One is the Triangle-lattice neighborhood, which is obtained from setting up cells on the node of the triangle lattice.
In the case of the Triangle-lattice neighborhood, there are four neighboring cells of the exit cell, i.e., $n_e=4$, thus a conflict between two, three, and four pedestrians occur.
The incident angles are calculated as in Fig. \ref{pre8_nn-ia} (b) since we see equilateral triangles by connecting the center of each cell.
Note that the Fig. \ref{pre8_nn-ia} (b) is just a schematic view, and the each cell is depicted as approximated size.
The situation as in Fig. \ref{pre8_nn-ia} (b) is realistic for the congested evacuation since pedestrians often stand between the two former pedestrians to see their way clear between the two heads.
The Triangle-lattice neighborhood also reproduces the hexagonal close packing of pedestrians as the two dimensional granular flow.
The other is the Moore neighborhood as in Fig. \ref{pre8_nn-ia} (c).
Pedestrians have eight choices when they move and there are five neighboring cells of the exit cell, i.e., $n_e=5$, thus there is a possibility of occurring a conflict between two to five pedestrians at the exit.
The incident angles are as in Fig. \ref{pre8_nn-ia} (c).
The difference between the Neumann neighborhood and the Moore neighborhood in the FF model is studied in Ref. \cite{moore1}.

When we consider the pedestrian dynamics in the low density case, where there are a few pedestrians, the number of neighboring cells does not greatly affect on the number of conflicts, however, it becomes significantly different according to the neighborhood type in the high density case.
Since we study an evacuation situation, the density at the exit is always high, thus we verify which neighborhood is the most suitable for analyzing evacuation by our experiments in Sec. \ref{EXP1}.

\SEC{Analytical Expression for the Average Pedestrian Outflow using the Cluster Approximation}{AOCA}

\subsection{Approximate Transition Probability}

We study the theoretical expression for the average pedestrian outflow through an exit in an extremely congested situation for arbitrary kinds of neighborhoods by cluster approximation in the FF model.
We consider a room with an exit with one-cell, and focus on the exit cell and its neighboring cells $1$ to $n_e$ in Fig. \ref{pre10_cluster}.
The number of the neighboring cells of the exit is described as arbitrary $n_e\in\textbf{N}$ to obtain the expression of the outflow for arbitrary kinds of neighborhoods.
The case $n_e=3$, $n_e=4$, and $n_e=5$ correspond to the case Neumann neighborhood, Triangle lattice neighborhood, and Moore neighborhood, respectively, as in Fig. \ref{pre8_nn-ia}.

We assume $k_s\rightarrow \infty$ since pedestrians near the exit try to evacuate in the shortest way.
Then, a pedestrian at neighboring cell $m\in[1,n_e]$, which we call pedestrian-$m$ in the following, has only two choices, i.e., moving to the exit cell with the probability $p(m \rightarrow \textbf{E})$ or stay at his/her cell $p(m \rightarrow m)$, which are described as
\EQN{ \label{pmEpmm}
\LPS{
p(m \rightarrow \textbf{E}) &= (p_{ij})_{\beta}|_{(i,j)=\textbf{E}} \rightarrow \beta, \\
p(m \rightarrow m) &= (p_{0,0})_{\beta} \hspace{0.94cm} \rightarrow 1-\beta, \hspace{0.2cm} (k_s \rightarrow \infty), \\
}}
respectively.

A pedestrian who is at the exit cell also has only two choices, i.e., getting out with the probability $p(\textbf{E} \rightarrow \rm{out})$ or stay at the exit cell with the probability $p(\textbf{E} \rightarrow \textbf{E})$, which are described as follows:
\EQN{ \label{alpha}
\LPS{
p(\textbf{E} \rightarrow \rm{out}) &= \alpha \tau(\theta_{e,m}), \\
p(\textbf{E} \rightarrow \textbf{E}) \hspace{0.25cm} &= 1 - \alpha \tau(\theta_{e,m}),
}}
where $\alpha$ is the probability of getting out from the room, which does not include the effect of turning (Sec. \ref{update}), and $\theta_{e,m}$ is the angle of incidence when pedestrians move from the cell $m$ .
We denote $\theta_{e,m}$ as $\theta_m$ in the following for simplicity as in Fig. \ref{pre10_cluster}.

$p(\textbf{E} \rightarrow \rm{out})$ represents both turning at the exit cell and getting out from the room (Fig \ref{pre9_turnseq}).
The turning function (\ref{tau}) is introduced to transition probability of the pedestrians at the exit cell since they need to change their headings quickly to face toward the exit (Fig. \ref{pre9_turnseq} (b)$\rightarrow$(c)), while we neglect the effect of turning at the neighboring cells of the exit for simplicity since we assume that all pedestrians at the neighboring cells are directed to the exit cell in the evacuation situation.




\FIGT{\EPSF{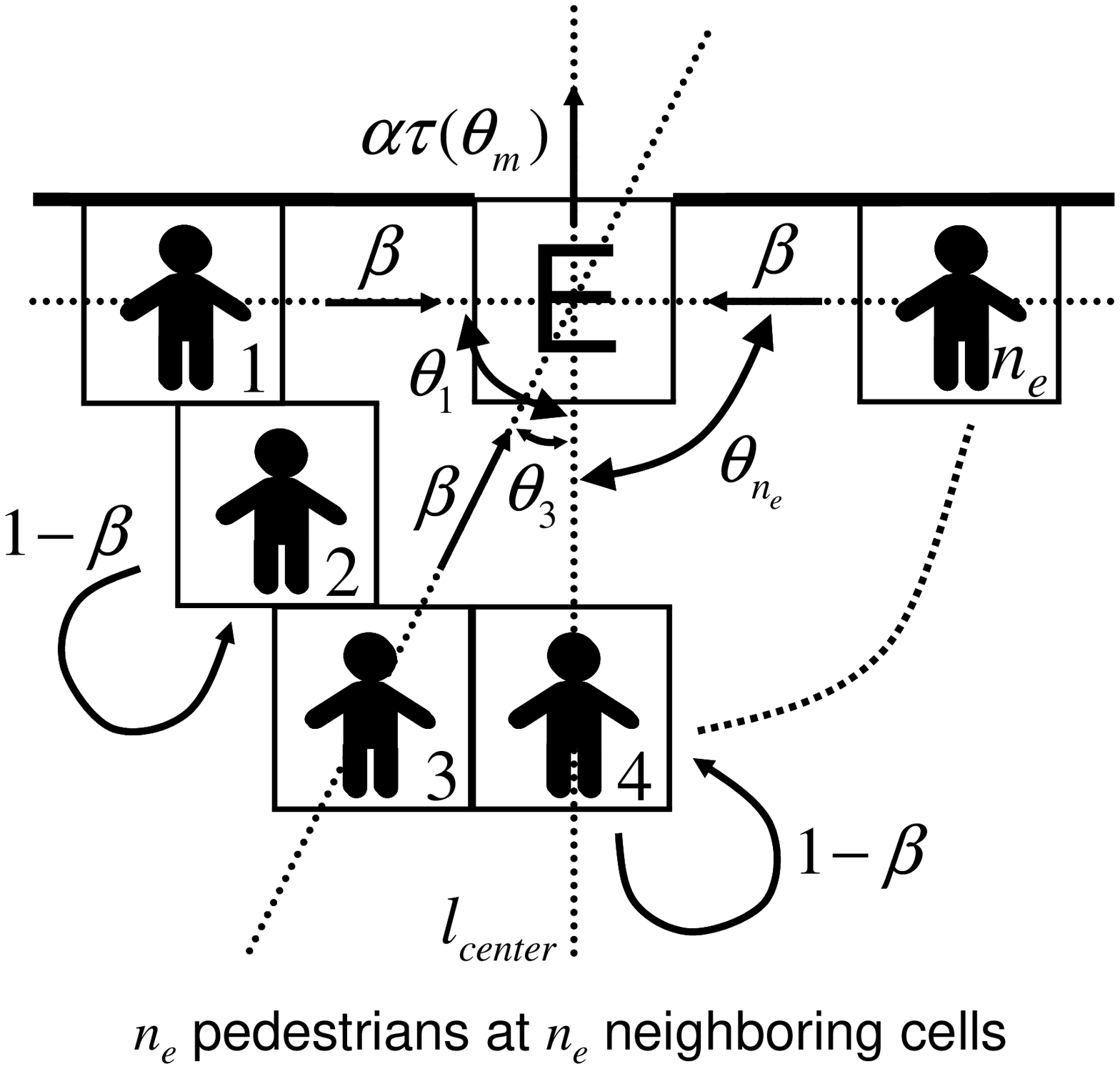}{
Schematic view of the cluster at the exit cell and its neighboring cells.
We consider the congested evacuation situation, thus all $n_e$ neighboring cells are occupied by pedestrians.
They try to move to the exit cell and conflict each other.
$\beta$ and $\alpha \tau(\theta_m)$ represent the transition probability to the exit cell (\ref{pmEpmm}), and the transition probability getting out from the exit cell (\ref{alpha}), respectively.
$\theta_m\ (m\in[1,n_e])$ is the incident angle when the pedestrian-$m$ moves to the exit cell.
}{height}{8.3}}

\SUB{Mathematical Formulation of \\ the Pedestrian Outflow}{AOCA-math}

Each neighboring cells of the exit has two states, i.e., occupied by a pedestrian and vacant, and the exit cell has $n_e+1$ states since we consider the headings of the pedestrians to introduce the turning effect.
Therefore, there are $(n_e+1)2^{n_e}$ states when we focus on the cluster at the exit cell and its neighboring cells.
We assume that pedestrians come into the neighboring cells of the exit with probability 1 since we consider the congested evacuation.
Then, the number of the states of the cluster reduces to $2n_e+1$.
Figure \ref{pre11_exit-state} shows the reduced states of the cluster in the Neumann neighborhood case $(n_e=3)$.
There is a state that the exit cell is vacant, i.e., (S0), and there are two states that the exit cell is occupied by pedestrian-$m$, i.e., (Sm-A) and (Sm-B), for each $m(=1,2,3)$ in the Fig. \ref{pre11_exit-state}.

\FIGT{\EPSF{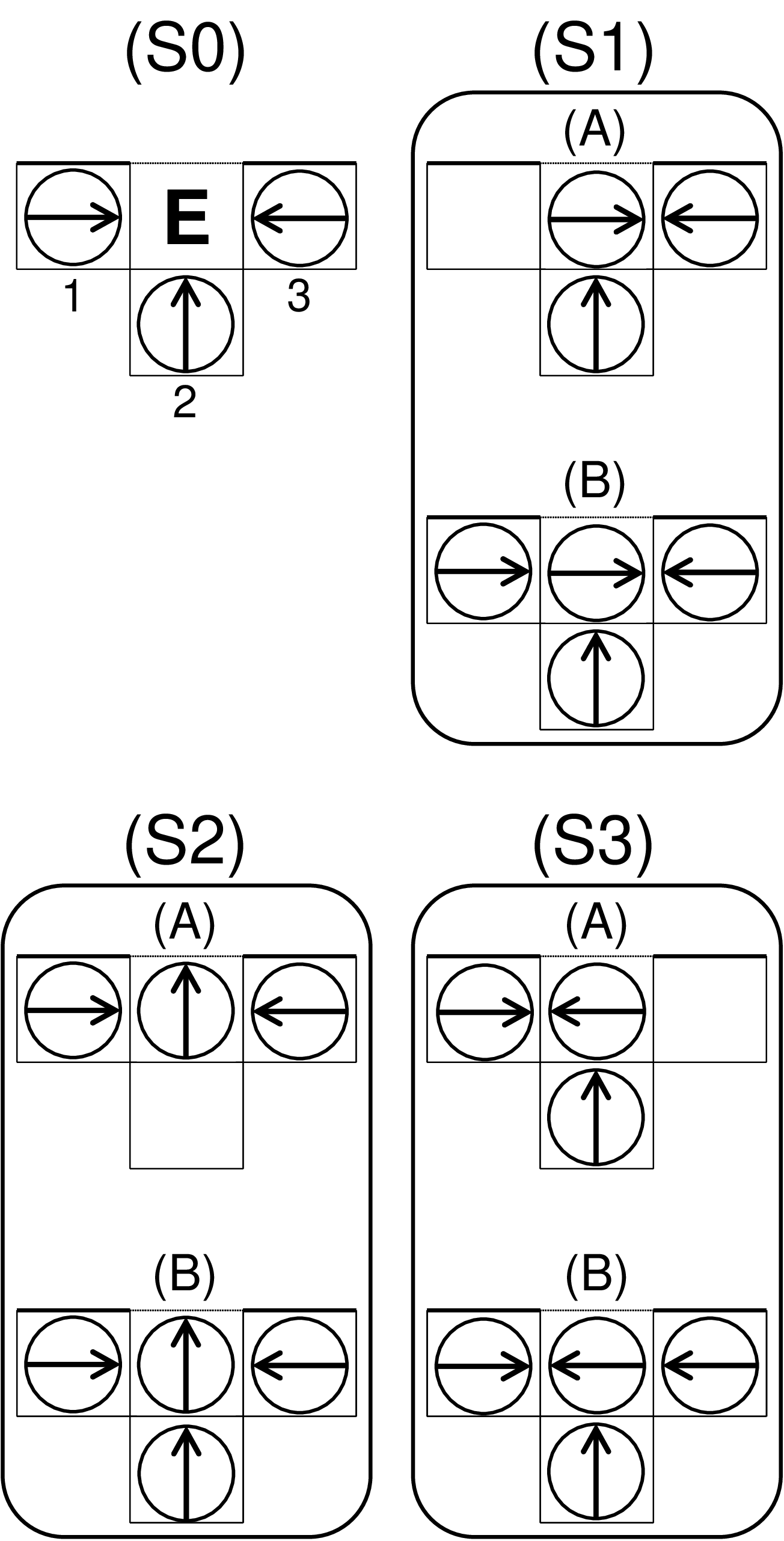}{
Schematic view of $7(=1+2\times3)$ states of the cluster in the Neumann neighborhood case $(n_e=3)$.
(S0) represents the state that the exit cell is vacant, and (S$m$) represents the state that the exit cell is occupied by pedestrian-$m$ $(m=1,2,3)$.
The pedestrian-1, 2, and 3 are circled $\rightarrow$, circled $\uparrow$, and circled $\leftarrow $, in the figure, respectively. 
The States (S$m$-A) and (S$m$-B) in each (S$m$) are combined when we calculate the pedestrian outflow. 
}{height}{14}}

We need the probability distribution of the exit-cell state to calculate the pedestrian outflow through the exit; however, do not need that of neighboring cells.
Thus, we combine the state (S$m$-A) with (S$m$-B) and consider the state transition among $n_e+1$ exit-cell states.
Figure \ref{pre12_sta-dia} is the state transition diagram of the exit cell in the Neumann neighborhood case $(n_e=3)$.
We find that the diagram is symmetrical with respect to the (S0).
$\alpha \tau(\theta_m)$ is the probability of getting out from the exit cell given by (\ref{alpha}). 
$r(n_e)$ represents the probability of one pedestrian move to the exit cell by solving a conflict when there are $n_e$ neighboring cells of the exit.
As we describe in Sec. \ref{collision}, the pedestrian which actually moves is chosen randomly with equal probability when the conflict is solved.
Thus, the transition probability of (S0) to (S$m$) is given by $r(n_e)/n_e$.
An explicit form of the $r(n_e)$ is obtained in the following. 

\FIGT{\EPSF{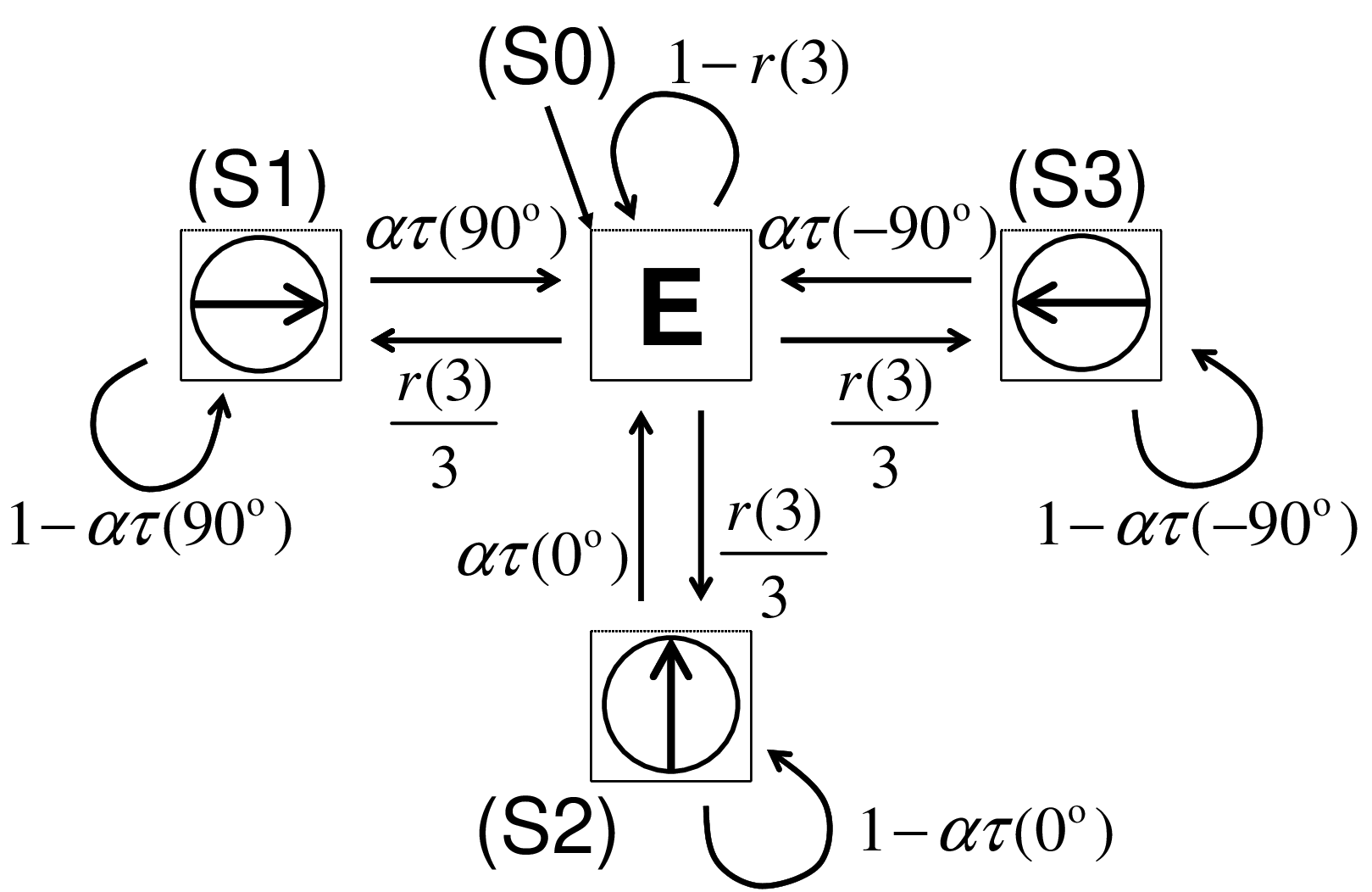}{
State transition diagram of the cluster in the Neumann neighborhood case $(n_e=3)$.
$r(n_e)$ is the probability that one of the pedestrian at the neighboring cells succeeds to move to the exit cell by solving conflicts (\ref{req}), and $\alpha \tau(\theta_m)$ is the probability of getting out from the exit cell (\ref{alpha}).
Since the diagram is symmetrical with respect to the (S0), we successfully calculate the average pedestrian outflow with arbitrary number of the neighboring cells $n_e$.
}{height}{5.5}}

Now, we start to calculate the average pedestrian outflow.
First, the probability of $k_e\in\textbf{N}$ pedestrians trying to move to the exit cell at the same time is described as
\EQ{
b(k_e) = \left( \AL {n_e \\ k_e } \right) \beta^{k_e} (1-\beta)^{n_e-k_e},
}
by using (\ref{pmEpmm}).
Next, $r(n_e)$, which is the probability of one pedestrian succeeds to move to the exit cell, is obtained as
\EQ{ \label{req}
r(n_e) = \sum_{k_e=1}^{n_e} \left[ \left\{ 1-\phi(k_e) \right\} b(k_e) \right],
}
where $\phi$ is the frictional parameter or function given by (\ref{phim}) or (\ref{phiz}), respectively.
We define $\pi_t(0)$ as the probability that the cluster is at the state (S0), i.e., a pedestrian is not at the exit cell at time step $t$ and $\pi_t(m)\ (m\in[1,n_e])$ as the probability that the cluster is at the state (S$m$), i.e., a pedestrian who was at the neighboring cell $m$ is at the exit cell.
Then, the master equation of the cluster is described as follows:
\EQ{ \SP{ \label{joutaiseni}
\begin{bmatrix}
\pi_{t+1}(0) \\ \pi_{t+1}(1) \\ \vdots \\ \pi_{t+1}(n_e)
\end{bmatrix}
= \hspace{5.5cm} \\
\begin{bmatrix}
1-r(n_e)	& \alpha \tau(\theta_1) 		& \ldots	& \alpha \tau(\theta_{n_e}) \\
r(n_e)/n_e	& 1-\alpha \tau(\theta_1) 	& 				& \bigzerou 	\\
\vdots	 											& 				& \ddots \\
r(n_e)/n_e	& \bigzerol	 					& 				& 1-\alpha \tau(\theta_{n_e}) 
\end{bmatrix}
\begin{bmatrix}
\pi_{t}(0) \\ \pi_{t}(1) \\ \vdots \\ \pi_{t}(n_e)
\end{bmatrix},
}}
where $\alpha \tau(\theta_e)$ represents the probability of getting out from the exit (\ref{alpha}).
By using (\ref{joutaiseni}) with the normalization condition
\EQ{
\sum_{m=0}^{n_e} \pi_t(m) = 1,
}
we obtain the stationary $(t \rightarrow \infty)$ solution
\EQ{ \LPS{
\pi_{\infty}(0) &= \left[ 1+\frac{r(n_e)}{n_e \alpha} \sum_{\acute{m}=1}^{n_e} \frac{1}{\tau(\theta_{\acute{m}})} \right]^{-1}, \\
\pi_{\infty}(m) &= \frac{r_(n_e)}{n_e\alpha \tau(\theta_m)} \pi_{\infty}(0) \hspace{0.6cm} (m\in[1,n_e]).
}}
Thus, the number of pedestrians who can evacuate from an exit with one cell's width per one time step, i.e., the average pedestrian outflow through the exit, is described as
\EQL{flow}{ \SP{
\LRA{q_{\rm{theo}}(n_e,\theta_1,\ldots,\theta_{n_e})} &= \sum_{m=1}^{n_e} \alpha \tau(\theta_m) \pi_{\infty}(m) \\
&= r(n_e) \pi_{\infty}(0) \\
&= \left[ \frac{1}{r(n_e)} + \frac{1}{n_e \alpha} \sum_{\acute{m}=1}^{n_e} \frac{1}{\tau(\theta_{\acute{m}})} \right]^{-1},
}}
where $\LRA{x}$ represents the sample average of $x$.

\SUB{Four Formulations of the Pedestrian Outflow}{reduc}

There are two choices to represent conflicts, i.e., using frictional parameter $\phi_\mu$ or frictional function $\phi_\zeta$.
We also consider whether the turning effect is included or not for each outflow.
Therefore, we obtain four kinds of the theoretical pedestrian outflows as follows:
\begin{itemize}

\item $\LRA{q_{\mu 0}} = \LRA{q_{\rm{theo}}}|_{\phi=\phi_\mu,\eta=0}$ \\
The simplest formulation of the pedestrian outflow, which includes the effect of conflict by the frictional parameter and neglects the turning effect.
The results obtained in Ref. \cite{dy1pre} is recovered by substituting $n_e=1,2,3$ as follows:
\EQ{
\LRA{q_{\mu 0}(1)} = \frac{\alpha \beta}{\alpha + \beta}, \hspace{4.3cm}
}
\EQ{
\LRA{q_{\mu 0}(2)} = \left[ 1- \frac{\alpha}{\alpha + 2\beta - (1+\mu) \beta^2} \right],
\hspace{1.28cm}
}
\EQN{
\LRA{q_{\mu 0}(3)} = \hspace{5.65cm} \nonumber \\
\alpha \left[ 1- \frac{\alpha}{\alpha_0 + 3\beta - 3(1+\mu) \beta^2 + (1+2\mu)\beta^3} \right]. 
}

\item $\LRA{q_{\zeta 0}} = \LRA{q_{\rm{theo}}}|_{\phi=\phi_\zeta,\eta=0}$ \\
The pedestrian outflow which includes the effect of conflicts by the frictional function and does not consider the effect of turning.
This formulation is obtained in Ref. \cite{dy3ped} as
\EQ{ \SP{
\LRA{q_{\zeta 0}(n_e)}
&= \left[ \frac{1}{r(n_e)} + \frac{1}{n_e} \sum_{\acute{m}=1}^{n_e} \frac{1}{\alpha} \right]^{-1} \\
&= \frac{\alpha r(n_e)}{\alpha + r(n_e)}
}}

\item $\LRA{q_{\mu \eta}} = \LRA{q_{\rm{theo}}}|_{\phi=\phi_\mu}$ \\
In this pedestrian outflow formulation, conflicts and turning are considered by the frictional parameter and the turning function, respectively. 

\item $\LRA{q_{\zeta \eta}} = \LRA{q_{\rm{theo}}}|_{\phi=\phi_\zeta}$ \\
The pedestrian outflow, which includes the effect of conflicts and turning by the two new functions introduced in this paper, i.e., the frictional function $\phi_{\zeta}$ and the turning function $\tau$. \\

\end{itemize}

In the Sec. \ref{EXP1} and \ref{EXP2}, we verify which formulation is the most suitable for calculating the realistic pedestrian outflow.

\FIGT{\EPSF{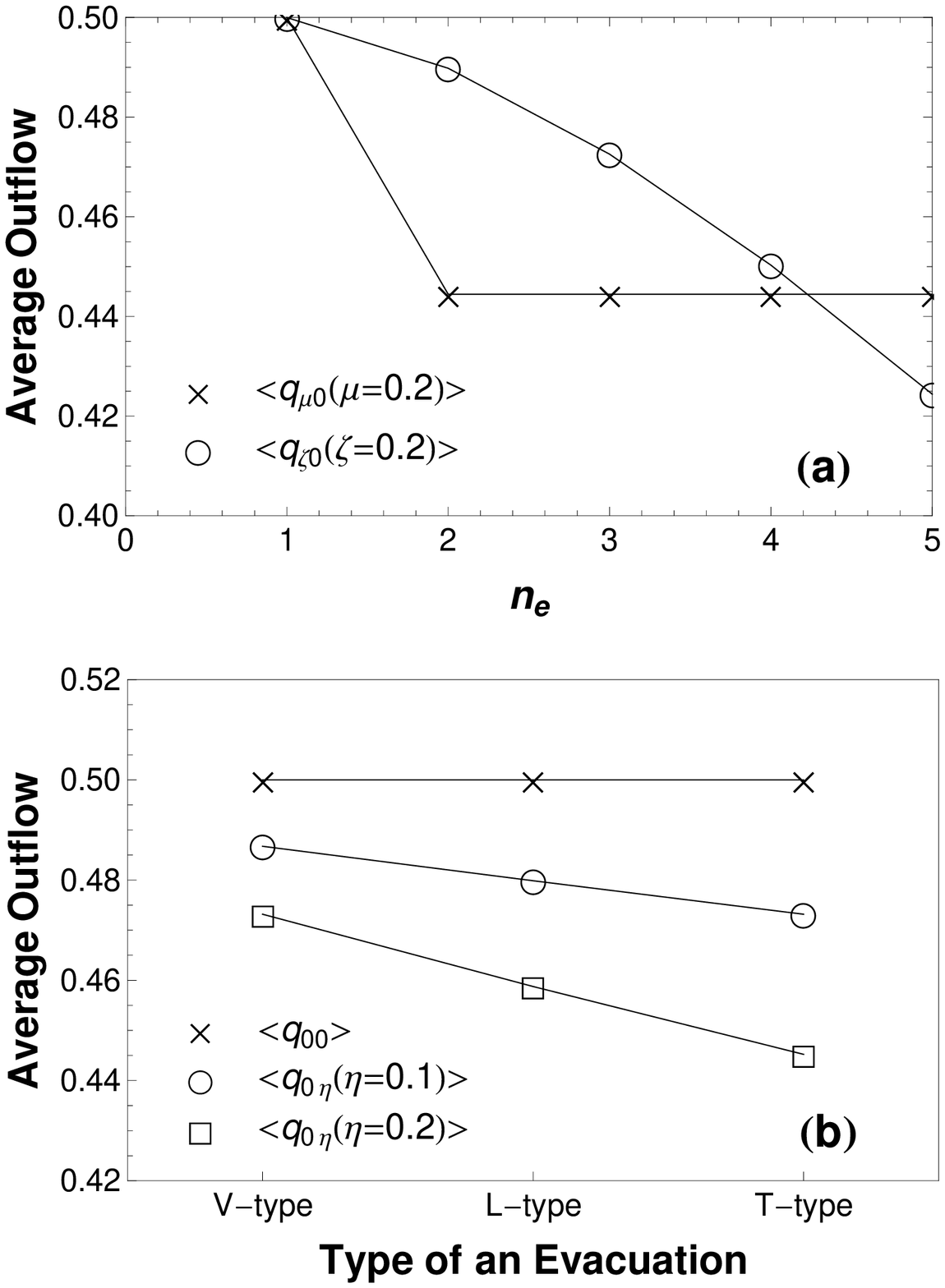}{
(a) Average pedestrian outflows $\LRA{q_{\mu 0}}$ and $\LRA{q_{\zeta 0}}$ as functions of the number of the neighboring cells of the exit $n_e$ in the case $\beta=1,\ \alpha=1$.
$\LRA{q_{\mu 0}}$ does not change in $n_e\geq2$, while $\LRA{q_{\zeta 0}}$ keeps decreasing in the all region in $n_e\geq1$.
(b) Average pedestrian outflows $\LRA{q_{0 0}}$ and $\LRA{q_{0 \eta}}$ for different evacuation types in the case $\beta=1,\ \alpha=1$.
The explanation of the three evacuation types are described in Fig. \ref{pre14_vlt}.
We see that $\LRA{q_{0 0}}$ does not change according to the three evacuation types, whereas $\LRA{q_{0 \eta}}$ does change.
The slope becomes sharp when $\eta$ increases.
}{height}{11.5}}

\FIGT{\EPSF{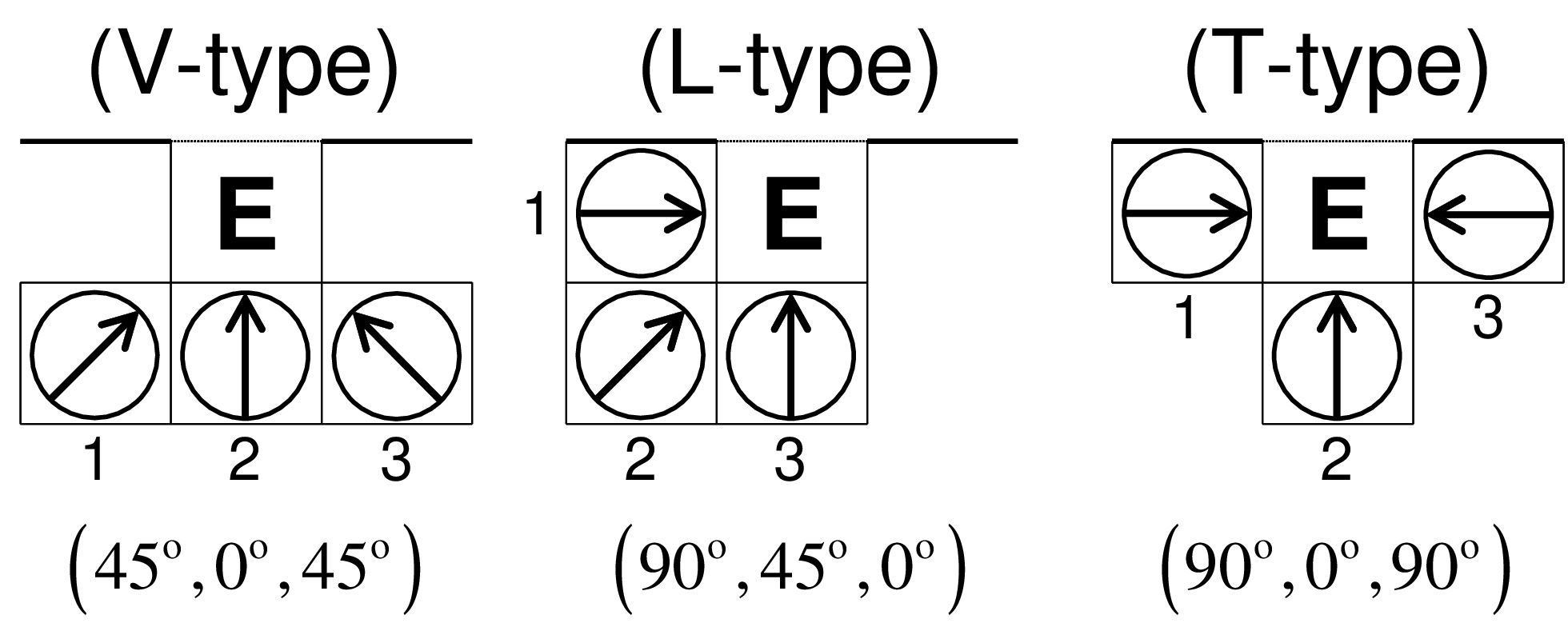}{
Schematic views of the three evacuation types in the case $(n_e = 3)$.
The neighboring cells are denoted as 1, 2, and 3.
The figures described below the schematic views are the incident angles $(\theta_1,\theta_2,\theta_3)$. 
The names of the evacuation types are decided from the directions from which pedestrians come into the exit cell.
}{height}{3.5}}

\FIGWT{\EPSF{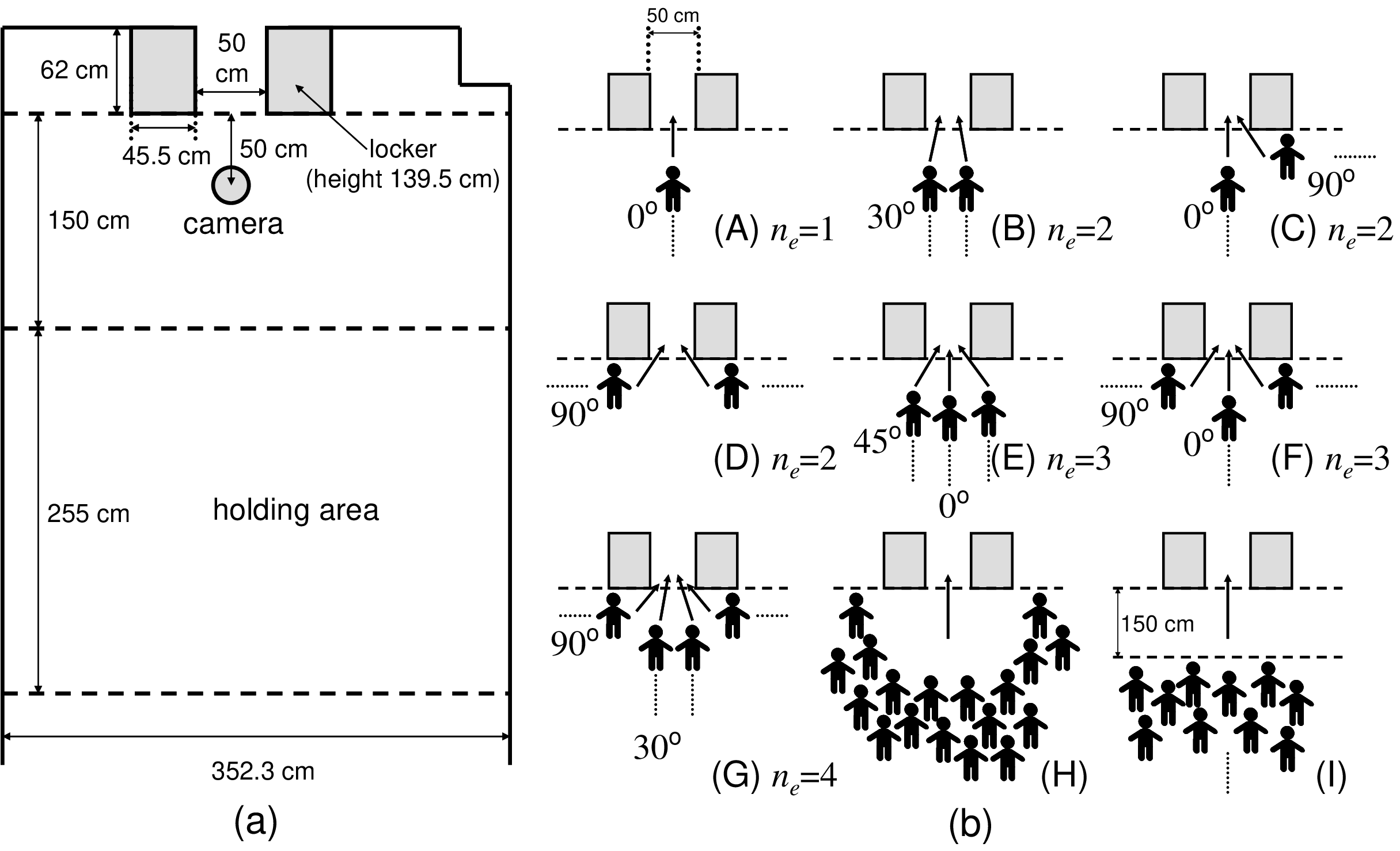}{
(a) Schematic view of the room used in the experiments. 
The width of the door is 50 cm with a locker at both sides.
(b) Schematic views of nine conditions of the experiments.
In the case from (A) to (G), pedestrians evacuate in lines.
They are forbidden to pass the former pedestrian.
The figures described by the silhouettes of pedestrians are the approximated incident angles used in theoretical calculation.
The participants of the experiments are not exactly go to the exit along the angle of incidence described in the figures.
The way of the evacuation is same in (H) and (I), however, the initial conditions are different as in the figures.
}{height}{10.8}}

\subsection{Comparison of the Theoretical Outflow}

Figure \ref{pre13_gqg00h} (a) shows the average pedestrian outflow $\LRA{q_{\mu 0}}$ and $\LRA{q_{\zeta 0}}$, which are outflows that includes the effect of conflicts but does not include the effect of turning.
We see the difference between the effect of the frictional parameter $\phi_\mu$ and the frictional function $\phi_\zeta$ from the figure.
When we use frictional parameter, the outflow does not change as $n_e$, which is the number of neighboring cells of the exit cell, increases in $n_e \geq 2$, while it continues to decrease when we adopt the frictional function.
Since it is natural to consider that the strength of the clogging and sticking become stronger according to the increase of the number of pedestrians involved in the conflicts, $\LRA{q_{\zeta 0}}$ seems more realistic than $\LRA{q_{\mu 0}}$.
We obtain the decrease of the outflow in $n_e \geq 2$ in the urgent evacuation situation $(\beta=1)$ by the frictional function for the first time in this paper.

Figure \ref{pre13_gqg00h} (b) shows the average pedestrian outflow $\LRA{q_{0 0}}$ and $\LRA{q_{0 \eta}}$.
In these cases, the effect of conflicts is neglected by substituting 0 for $\mu$ or $\zeta$, and we see how the effect of turning influences on the outflow.
The schematic views of the three evacuation types, which are \textit{V-type}, \textit{L-type}, and \textit{T-type} are depicted in Fig. \ref{pre14_vlt}.
$n_e=3$ in all three types, however, the incident angles of the three pedestrians $\theta_1$, $\theta_2$, and $\theta_3$ are different.
When we neglect the turning effect ($\LRA{q_{0 0}}$), there is no difference of the outflow among the three evacuation types. 
However, when we consider the effect of turning by the turning function ($\LRA{q_{0 \eta}}$), the outflow decreases as the incident angles increase.
The difference of the pedestrian outflow according to the difference of the way of pedestrians gathering around the exit is observed since we have introduced the significant factor in the evacuation through the narrow exit, i.e., the decrease in walking speed by turning, by the turning function.

\SEC{Experiment and Analysis 1 Evacuation in lines}{EXP1}

\subsection{Experiment} \label{EXP1-exp}
We have performed the evacuation experiments to verify the relation between the two new factors and the average pedestrian outflow.
The schematic view of the room is described as in Fig. \ref{pre15_ex1} (a).
The width of the exit is 50 cm.
There are eighteen participants for the experiment, who are all men.
The experiments started when we clapped our hands and finished when the all pedestrians evacuated from the room.
Nine kinds of conditions are put into practice (Fig. \ref{pre15_ex1} (b)).
In the case (H) and (I) pedestrians could move as they want after the evacuation started, however, in the other cases pedestrians had to follow the former pedestrian, i.e., they were prohibited from putting the queues into disorder.
Each experiment was performed two or three times, which is sufficient for pedestrian dynamics experiments.
When we perform the same experiments many times, pedestrians memorize and study the situation, which greatly affects the experimental results.

The outflows are calculated as follows:
\EQ{ \label{qexp}
\LRA{q_{\rm{exp}}} = \frac{j-i}{w (t_{j} - t_{i})},
}
where $i$ and $j$ are the orders of the first and the last pedestrian used to calculate the pedestrian outflows, and $t_{i}$ and $t_{j}$ are the evacuation times of the $i$ th pedestrian and the $j$ th one, respectively.
$w$ represents the width of the exit.

We decide $i=4,\ j=18$ for the case (A) to (D) and $i=1,\ j=15$ for the case (E) to (H).
In the former cases, pedestrians move very fast at the starting time since there is no one ahead of the first pedestrian and the density is very low.
Thus, we need to wait for the stable state.
By contrast, in the latter cases, the evacuation becomes saturated in the stable state immediately since there are interactions among the pedestrians from the start.
However, in the end of the evacuation, the number of pedestrians decreases and the interaction among more than two pedestrians is hardly observed.
Consequently, the situations become totally different from the stable states.
Therefore, we use the evacuation time of first and fifteenth pedestrian for the calculation.
In the case (I), it takes some time to achieve the stable state since pedestrians are not gathering around the exit at the beginning of the evacuation.
In the end of the evacuation, the same phenomenon as the experiment (E) to (H) is observed.
Thus, we determine $i=4$ and $j=15$ for the case (I).

\TABLET{
\NAME{exp-result1}{
The pedestrian outflows of the experiments.  
``Case" in the table is corresponding to the case in Fig. \ref{pre15_ex1} (b).
$n_e$ is the number pedestrian lines in the experiments, which represents the number of the neighboring cells in the theoretical calculation. 
$\bm{\theta_e}=(\theta_1,\ldots,\theta_{n_e})$ is the approximated incident angles of the pedestrians.
The pedestrian outflows $\langle q_{\rm{exp}} \rangle $ [persons/(m$\cdot$ sec)] in the table are the average of the $N$ experiments, where $N$ is the number of the experiments described in the fifth column of the table. 
$i$ and $j$ are the orders of the first and last pedestrians which used to calculate the experimental outflow by (\ref{qexp}). 
}
\begin{tabular}{|c||c|c|c|c|c|c|}
\hline
\hspace{0.1cm} Case \hspace{0.1cm} & \hspace{0.1cm} $n_e$	\hspace{0.1cm} & $\bm{\theta_e}$ & \hspace{0.1cm} $\langle q_{\rm{exp}} \rangle$  \hspace{0.1cm} &  \hspace{0.1cm} $N$  \hspace{0.1cm}  & \hspace{0.1cm} $i$ \hspace{0.1cm} & \hspace{0.1cm} $j$ \hspace{0.1cm} \\
\hline \hline
(A)	& 1			& $(0^{\circ})$					& 2.62	& 3  & 4 & 18\\
\hline
(B)	& 2			& $(30^{\circ},30^{\circ})$	& 2.81	& 3  & 4 & 18\\
\hline
(C)	& 2			& $(0^{\circ},90^{\circ})$	& 2.69	& 3  & 4 & 18 \\
\hline
(D)	& 2			& $(90^{\circ},90^{\circ})$	& 2.59	& 3  & 4 & 18 \\
\hline
(E)	& 3			& $(45^{\circ},0^{\circ},45^{\circ})$	& 2.69	& 2 & 1 & 15 \\
\hline
(F)	& 3			& $(90^{\circ},0^{\circ},90^{\circ})$	& 2.59	& 3 & 1 & 15 \\
\hline
(G)	& 4			& $(90^{\circ},30^{\circ},30^{\circ},90^{\circ})$	& 2.51	& 2 & 1 & 15 \\
\hline
(H)	& -			& - & 2.51	& 3 & 1 & 15 \\
\hline
(I)	& -			& - & 2.53	& 2 & 4 & 15 \\
\hline
\end{tabular}
}

The results of the experiments are described in Tab. \ref{exp-result1}.
There are four remarkable points in the table.
First, we see that the pedestrian outflow decreases in $n_e\geq 2$ by comparing the case (B), (E), and (G).
This indicates that when the number of the pedestrian-lines increases, the outflow decreases since the possibility of the conflicts increases as in Fig. \ref{pre5_mu012}.
Second, we observe that the outflow decreases with the increase of the components of $\bm{\theta_e}$ when $n_e$ is constant by comparing (B), (C), and (D) or (E) and (F).
We have verified that turning at the exit does decrease the outflow.
Third, the maximum outflow is attained at $n_e=2$.
See the Fig. \ref{pre16_ex1-B}, which is a snapshot of the experiment (B).
We see that pedestrians in the left line and the right line go through the exit one after the other.
This phenomenon is called zipper effect \cite{ex_bottleneck2}, when it occurs, the outflow increases since the number of conflicts between pedestrians decrease dramatically.
In the traffic flow, the \textit{alternative configuration}, which leads to the smooth merging as the pedestrian evacuation in the two-line case, is achieved by the \textit{compartment line} \cite{rn1arxiv}.
By contrast, pedestrians are intelligent enough to avoid conflicts by their selves.
They do not need any help to achieve the alternative configuration.
Forth, the pedestrian outflow in the case (G) $(n_e=4)$, (H), and (I) are similarly smaller than the other cases.
This indicates that there are approximately four pedestrians at the exit in the normal evacuation situation when the width of the exit is 50 cm.
Consequently, we have found that the Triangle-lattice neighborhood is the most suitable for the normal evacuation.

\FIGT{\EPSF{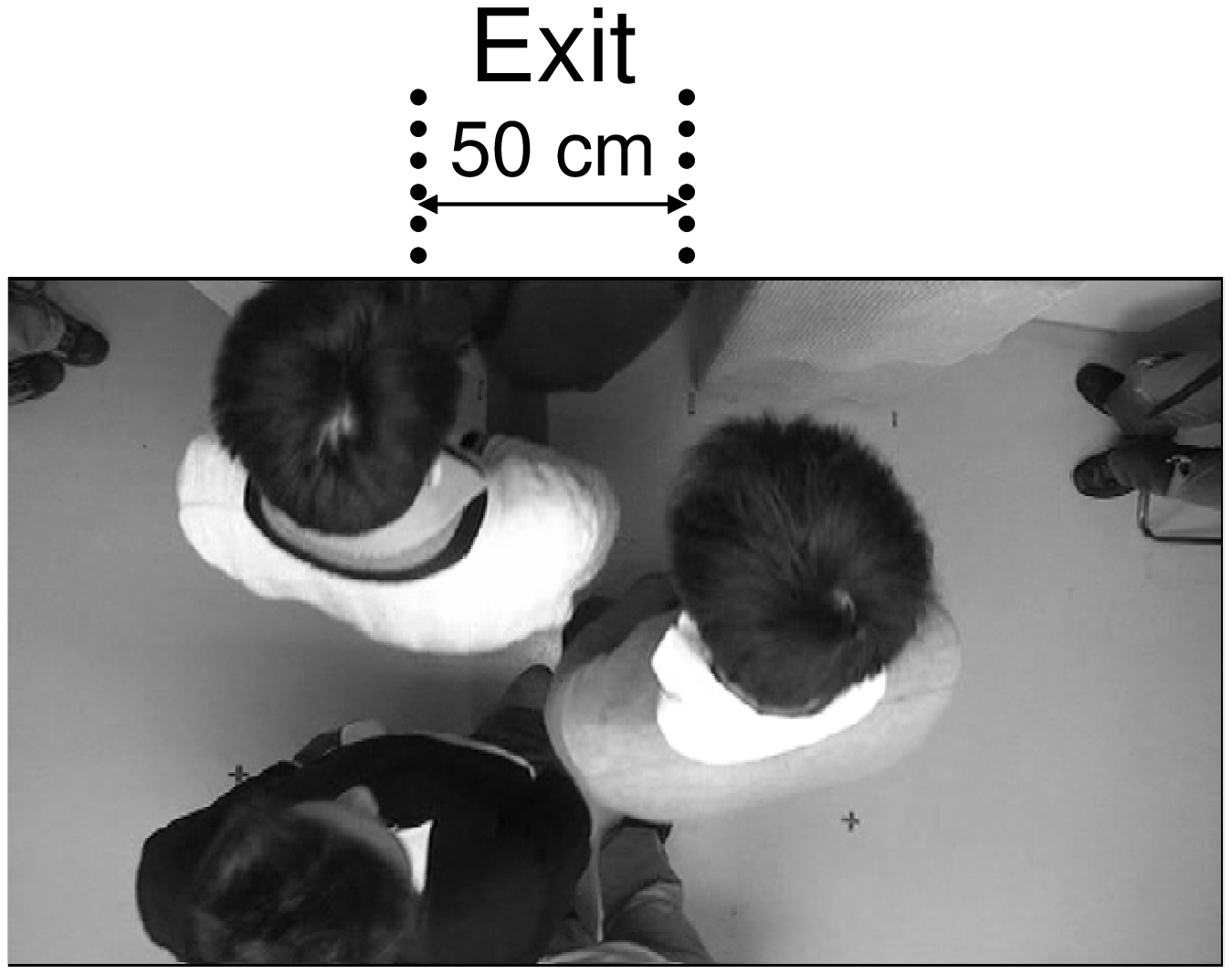}{
Snapshot of the experiment (B) $(n_e=2)$.
The pedestrians in the left line and the right line go through the exit one after the other without any instructions.
}{height}{6}}

\subsection{Theoretical Analysis} \label{EXP1-theo}

We calculate the average errors between the theoretical outflows and the experimental outflows as follows:
\EQL{err}{
\sqrt{
\frac{  \sum(\LRA{q_{\rm{theo}}} - \LRA{q_{\rm{exp}}})^2 }
{\rm{(Number\ of\ the\ kinds\ of\ the\ experiments)}}
},
}
and verify the effect of the frictional function and the turning function.
First, we define the length of the side of square cell $\Delta l$ and the time of one time step $\Delta t$ as 
\EQN{ \LP{
\Delta l = 0.5\ \rm{[m/cell]}, \\
\Delta t = 0.3\ \rm{[sec/step]}.
}}

Second, we determine $\alpha$ and $\beta$. In the situation (A) to (I), the pedestrians do not change their walking speeds instantaneously after they succeed to move to the exit cell.
Therefore, we assume $\alpha = \beta$.
Then, we calculate $\beta$ from the result of the experiment (A) as follows:
\EQN{\SP{
\LRA{q_{\rm{theo}}(n_e=1, \theta_1 =0^\circ)} &= 2.62 \times \Delta l \times \Delta t, \\ 
\beta &= 0.79.
}}
Note that when $n_e=1$, the frictional parameter, function, and turning function do not affect on the pedestrian outflow, so that we obtain common $\beta$ for the four kinds of outflows.

\TABLET{
\NAME{ana-result1}{
Parameters and the average errors for the experiment (A) to (I) from the method of least squares.
We see that the average errors calculated by (\ref{err}) are large in three cases, i.e., $\LRA{q_{\mu0}}$, $\LRA{q_{\zeta 0}}$, and $\LRA{q_{\mu \eta}}$.
However, if the both frictional and turning functions are applied to the outflow, i.e., $\LRA{q_{\zeta \eta}}$, the error becomes smaller than half of the other cases.
}
\begin{tabular}{|c||c|c|c|c|}
\hline
\hspace{0cm} $\LRA{q_{\rm{theo}}}$ \hspace{0cm} & \hspace{0.3cm} $\mu$ \hspace{0.3cm} & \hspace{0.3cm} $\zeta$ \hspace{0.3cm} & \hspace{0.3cm} $\eta$ \hspace{0.3cm} &  \hspace{0.1cm} average error  \hspace{0.1cm} \\
\hline \hline
$\langle q_{\mu 0} \rangle$	& 0.25		& -		& -			& 0.08	\\
\hline
$\langle q_{\zeta 0} \rangle$	& -		& 0.34		& -		& 0.08	\\
\hline
$\langle q_{\mu \eta} \rangle$	& 0.18		& -		& 0.07	& 0.07	\\
\hline
$\langle q_{\zeta \eta} \rangle$	& -		& 0.26		& 0.09	& 0.03	\\
\hline
\end{tabular}
}

Finally, we decide the parameters $\mu$ or $\zeta$, and $\eta$ for four theoretical average pedestrian outflow introduced in Sec. \ref{reduc} from the method of least squares.
The number of the pedestrian lines is assumed as $n_e=4$ in the case (H) and (I) since the outflow in the case (G) $(n_e=4)$ and in the case (H) and (I) are similarly small as we have discussed in the Sec. \ref{EXP1-exp}.
This assumption is also supported by the fact that pedestrians stands between two former pedestrians to see their way clear in the evacuation situation, which is achieved by the Triangle-lattice neighborhood $(n_e=4)$.

Table \ref{ana-result1} shows the parameters calculated by the method of least squares and the average errors for each outflow.
The errors of $\LRA{q_{\mu 0}}$, $\LRA{q_{\zeta 0}}$, and $\LRA{q_{\zeta \eta}}$ are larger than 0.05.
This means that only considering one factor, i.e., ``the difference of the strength of conflicts according to the number of involved pedestrians" or ``the decrease in walking speed by turning", is not sufficient to reproduce the realistic outflow.
By contrast, the error becomes smaller than 0.05, when both the frictional and turning functions are introduced.
This result verifies that the introduction of the frictional function and the turning function enables us to obtain the more realistic pedestrian outflows, theoretically.


The average pedestrian outflows of the theoretical analysis are compared with those of the experiments in Fig \ref{pre17_gE}.
We can clearly see the four remarkable points discussed in Sec. \ref{EXP1-exp}.
$\LRA{q_{\zeta \eta}}$, which includes the effect of both the frictional function and the turning function, agrees with the experimental results well except (B).
On the contrary, $\LRA{q_{\mu 0}}$ is not correspond to the experimental results well.
The effect of turning is not applied to $\LRA{q_{\mu 0}}$, thus its values are same if $n_e$ is same.
Moreover, it does not use the frictional function either, so that it is impossible to change its value for $n_e $ appropriately.
Therefore, we confirm that the frictional function and the turning function have an essential role to calculate the realistic pedestrian outflow again.

\FIGT{\EPSF{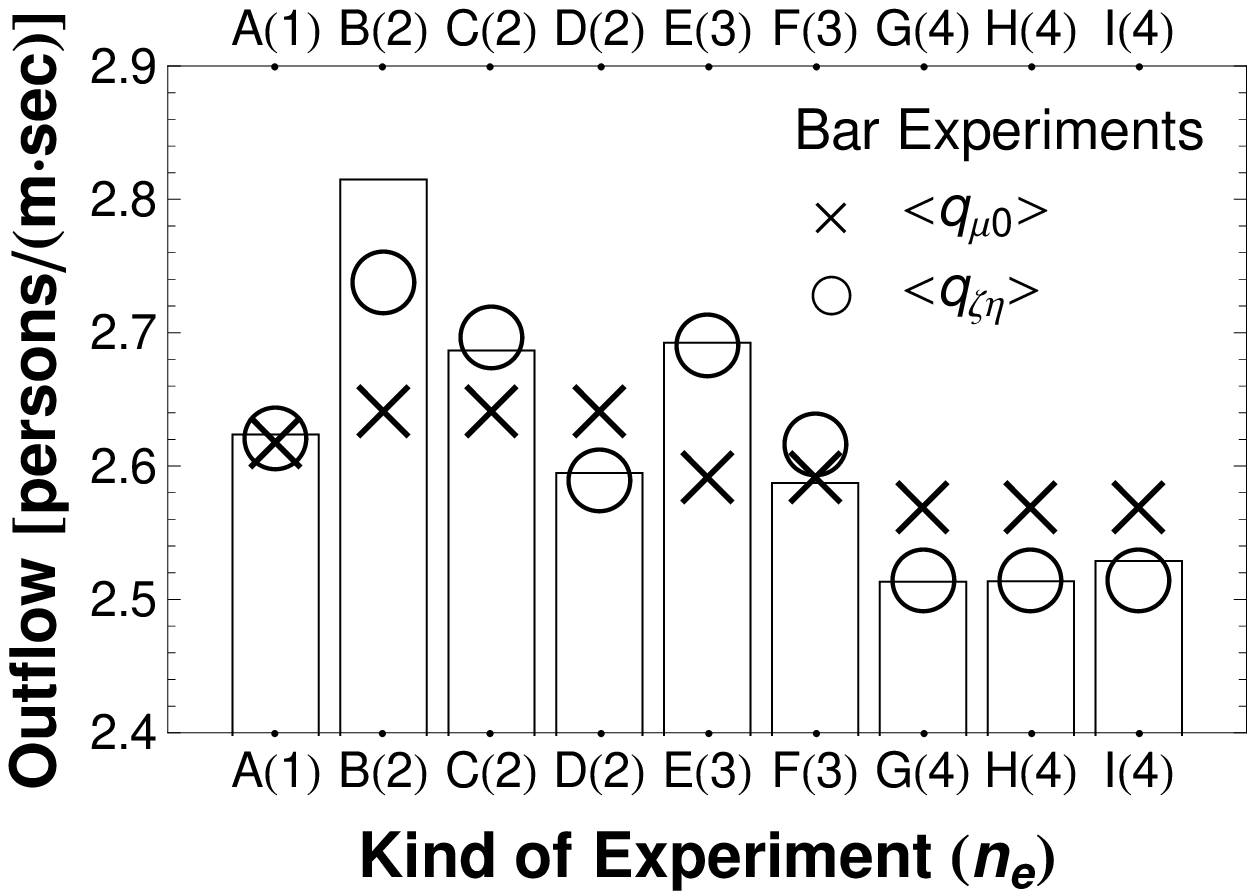}{
The pedestrian outflows from the experiments and the theoretical calculation.
The parameters in Tab. \ref{ana-result1} are used for the theoretical results.
We observe that $\LRA{q_{\zeta \eta}}$ agree with the experimental results well except (B), while $\LRA{q_{\mu 0}}$ does not.
}{height}{5.8}}

The outflow of the experiment (B) is extremely large and the result of the theoretical calculation does not agree with it even if the two functions are introduced because of the zipper effect.
Since the pedestrian intelligence enables them to achieve the smooth merging, it should be modeled to reproduce the zipper effect \cite{dy_preparation}.
However, when we consider the normal evacuation and the lined evacuation whose lines are more than two, the effect of the pedestrian intelligence is possible to neglect, and also the evacuation in two lines (B) is a special case; therefore, our model includes sufficient factors for analyzing usual evacuations.

\SEC{Experiment and Analysis 2 Evacuation through an Exit \break with an Obstacle}{EXP2}

\SUB{Experiment}{EXP2-exp}

We study the effect of an obstacle put in front of an exit in this section.
We did the evacuation experiment at the NHK TV studio in Japan.
Two large walls were set up in the studio, and we could adjust the width of the exit by moving them (Fig. \ref{pre18_ex2} (a)).
We decided it as 50 cm.
The participants of the experiment were fifty women, who were their thirties and forties.
We did three kinds of experiments as in Fig. \ref{pre18_ex2}.
The experiment (a) (Fig. \ref{pre18_ex2} (a)) was the evacuation in a line, which was the same kind of experiment as the experiment (A) in Fig. \ref{pre15_ex1} (b)-(A).
The experiment (b) (Fig. \ref{pre18_ex2} (b)) was the normal evacuation, whose initial condition was the same as the experiment (H) in Fig. \ref{pre18_ex2} (b)-(H).
The initial condition and the way of the evacuation in the experiment (c) (Fig. \ref{pre18_ex2} (c)) was the same as the experiment (b), however, the column, whose diameter was 20 cm, was put in front of the exit.

\FIGT{\EPSF{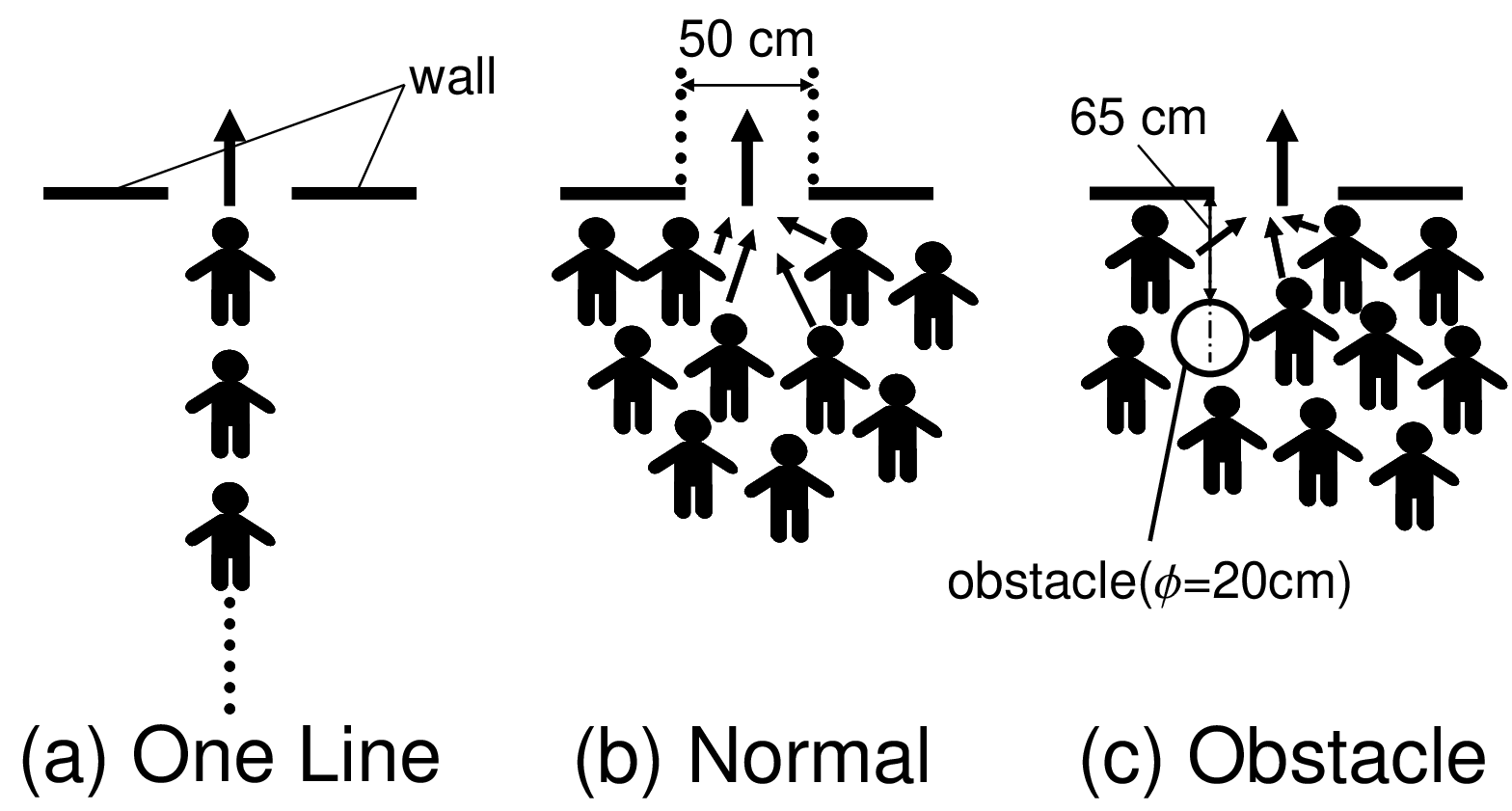}{
Schematic views of the experiments.
(a) ``Going through the exit in a line". (Same kind of experiment as Fig. \ref{pre15_ex1} (b)-(A).)
(b) ``Normal evacuation". (Same kind of experiment as Fig. \ref{pre15_ex1} (b)-(H).)
(c) ``Putting an obstacle in front of the exit".
The width of the exit is 50 cm and the number of the pedestrians is fifty.
We consider that the four pedestrians try to move to the exit at the same time in (b), while three pedestrians do in (c) since there is an obstacle.
}{height}{4.5}}

The pedestrian outflows of the three cases are described in Tab. \ref{exp-result2}.
We calculate them by using (\ref{qexp}), where $i$ and $j$ are determined as Tab. \ref{exp-result2} by considering the same reason as in Sec. \ref{EXP1-exp}.
The outflow in the case (a) is the largest, since there was no conflict.
Comparing the result of the experiment (b) and (c), we surprisingly find that the outflow of the experiment (c), i.e., the experiment putting an obstacle in front of the exit, is larger than that of the experiment (b), which is a normal evacuation.
The P value of the difference of the average pedestrian outflow between (b) and (c) is calculated as $p<0.01$, which verifies that the difference is significant.

\SUB{Theoretical Analysis}{EXP2-theo}

\TABLEB{
\NAME{exp-result2}{
The pedestrian outflows obtained from the experiments.
The data show that the outflow increases if we put an obstacle in front of the exit in a proper way.
We verify that there is a significant difference between the case (b) and (c) by calculating P value, which is obtained as $p<0.01$.
}
\begin{tabular}{|c||c|c|c|c|c|c|}
\hline
\hspace{0.02cm} Case \hspace{0.02cm} & \hspace{0.02cm} $n_e$	\hspace{0.02cm} & $\bm{\theta_e}$ & \hspace{0.02cm} $\langle q_{\rm{exp}} \rangle$  \hspace{0.02cm} &  \hspace{0.02cm} $N$  \hspace{0.02cm}  & \hspace{0.02cm} $i$ \hspace{0.02cm} & \hspace{0.02cm} $j$ \hspace{0.02cm} \\
\hline \hline
(a) One line	& 1			& $(0^{\circ})$					& 3.23	& 3  & 4 & 50 \\
\hline
(b) Normal \hspace{0.01cm}	& 4			& $(90^{\circ},30^{\circ},30^{\circ},90^{\circ})$	& 2.80	& 6  & 1 & 47 \\
\hline
(c) Obstacle	& 3			& $(90^{\circ},30^{\circ},90^{\circ})$	& 2.92	& 6  & 1 & 47 \\
\hline
\end{tabular}
}

We explain this phenomenon by our theoretical calculation.
Watching the video of the experiment, we see that the obstacle blocks the pedestrians moving to the exit.
In the Sec. \ref{EXP1-theo}, we have verified that there are approximately four pedestrians move to the exit at the same time, i.e., $n_e=4$, in the normal evacuation (Fig. \ref{pre18_ex2} (b)).
We assume that an obstacle blocks the pedestrians moving, so that the number of the pedestrians moving to the exit at the same time decreases by one.
Therefore, we have modeled the obstacle by blocking one neighboring cell as in Fig. \ref{pre19_oban} (a) and consider that $n_e=3$ in the experiment (c) (Fig. \ref{pre18_ex2} (c)).
This modeling is justified by the experimental result that 35 \% pedestrians went through the left side of the obstacle, whereas 65 \% of them went through the right side of it, that is, the ratio between the pedestrians going through the left side and right side of the obstacle is approximately $1:2$, which corresponds to the number of the left and right neighboring cells in regard to the obstacle cell in Fig. \ref{pre19_oban} (a).

Now, we obtain the parameter $\beta$ from the result of the experiment (a) as
\EQN{\SP{
\LRA{q(n_e=1, \theta _1=0^\circ)} &= 3.23 \times \Delta l \times \Delta t, \\ 
\beta &= 0.97.
}}
Then, we calculate $\mu$ or $\zeta$, $\eta$, and the average errors (\ref{err}) by the method of the least squares.

\TABLET{
\NAME{ana-result2}{
Parameters and the average errors for the experiment (a) to (c) from the method of least squares.
$\mu$, $\zeta$, and $\eta$ calculated by the least square method are described in the second, third, and forth columns, respectively.
In the fifth column, the average error is described.
We observe that there are errors order of $10^{-2}$ for $\LRA{q_{\mu 0}}$, $\LRA{q_{\zeta 0}}$, and $\LRA{q_{\mu \eta}}$, while $\LRA{q_{\zeta \eta}}$ agrees with the experimental data very well.
}
\begin{tabular}{|c||c|c|c|c|}
\hline
\hspace{0cm} $\LRA{q_{\rm{theo}}}$ \hspace{0cm} & \hspace{0.3cm} $\mu$ \hspace{0.3cm} & \hspace{0.3cm} $\zeta$ \hspace{0.3cm} & \hspace{0.3cm} $\eta$ \hspace{0.3cm} &  \hspace{0.1cm} average error  \hspace{0.1cm} \\
\hline \hline
$\LRA{q_{\mu 0}}$ & 0.23 & - & - & 0.05 \\
\hline
$\LRA{q_{\zeta 0}}$ & - & 0.27 & - & 0.04 \\
\hline
$\LRA{q_{\mu \eta}}$ & 0.23 & - & 0 & 0.05 \\
\hline
$\LRA{q_{\zeta \eta}}$ & - & 0.22 & 0.09 & 0 \\
\hline
\end{tabular}
}

The results are shown in Tab. \ref{ana-result2}.
We see that the average pedestrian outflows calculated by using the frictional function $\phi_\zeta$ and the turning function $\tau$, i.e., $\LRA{q_{\zeta \eta}}$, reproduces the experimental outflow very well since there is no error.
However, there are average errors whose orders are $10^{-2}$ for other three theoretical outflows, i.e., $\LRA{q_{\mu 0}}$, $\LRA{q_{\zeta 0}}$, and $\LRA{q_{\mu 0}}$.
The frictional parameter does not change the strength of the clogging against $n_e$; therefore, the difference of the outflow in the case (b) and (c) is not appropriately reproduced.
When we adopt the frictional parameter, we also find that the turning function cannot work effectively since $\eta=0$ in the case $\LRA{q_{\mu \eta}}$ (Tab. \ref{ana-result2}).
On the contrary, the frictional function can adjust the strength of clogging adequately, thus, the result of the experiment is obtained if it is used with the turning function. 
This result also verify the importance of the frictional function and justifies our assumption that the obstacle increases the pedestrian outflow since it decreases $n_e$, which is the number of pedestrians moving to the exit at the same time.
The turning function plays an important role to represent the position of the obstacle.
It enables us to distinguish which cell is blocked by the obstacle, the cell whose incident angle is $30^\circ$ or $90^\circ$.
If we do not use the turning function, the effect of the obstacle's position, which is studied in the next section, is not represented properly, so that the average error becomes large in $\LRA{q_{\zeta 0}}$.

Figure \ref{pre20_gN} shows the experimental and theoretical average pedestrian outflows.
The well correspondence between the experimental flows and $\LRA{q_{\zeta \eta}}$ is observed.
However, $\LRA{q_{\mu 0}}$ is almost same for the case (b) and (c), and does not agree with the experimental results.

\FIGT{\EPSF{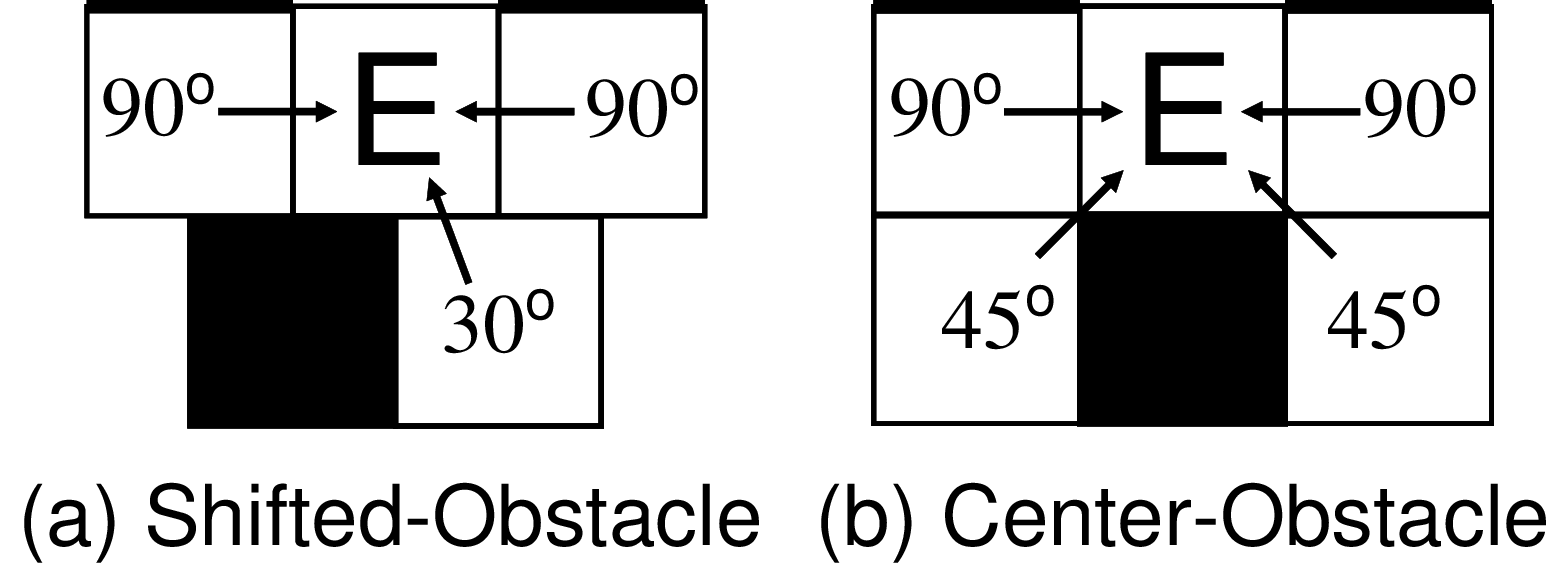}{
Schematic view of the exit cell and its neighbors when an obstacle is set up.
(a) The obstacle is shifted from the center.
(b) The obstacle is at the center.
We consider the incoming pedestrian flow from the diagonal neighboring cells of the exit cell since the obstacle is much smaller than the one cell.
}{height}{3}}

\SUB{Effect of the Position of the Obstacle}{EXP2-center}

We consider how the position of the obstacle affects on the pedestrian outflow in this section.
Fig. \ref{pre19_oban} (b) represents the case that the obstacle used in the experiment (c) in Fig. \ref{pre18_ex2} (c) is set at the center of the exit.
We assume that the pedestrians also move to the exit cell from the two diagonal neighboring cells since the diameter of the obstacle (20cm) is much smaller than the size of the cell (50 cm).
We calculate the pedestrian outflow in this case by using $\LRA{q_{\zeta \eta}}$ with the parameters in Tab. \ref{ana-result2}, which are $\beta=0.97$, $\zeta=0.22$, and $\eta=0.09$ as follows:
\EQ{ \SP{
\LRA{q_{\rm{center}}} &= \LRA{q(n_e=4,\bm{\theta_e}=(90^\circ, 45^\circ, 45^\circ, 90^\circ ) ) } \\
&= 2.78.
}}
This value is not only smaller than the outflow of the experiment (c), which is a evacuation with an shifted obstacle, but also the outflow of the experiment (b), which is a normal evacuation.
The result indicates that the obstacle does not always increase the pedestrian outflow.
If the position of the obstacle is not adequate, it does not block one pedestrian moving to the exit cell; therefore, the outflow is not improved.
Moreover, if the obstacle is at the center of the exit, pedestrians whose incident angles are $0^\circ$, i.e., who are going to go through the exit straightly, have to detour the obstacle, so that their walking speeds decrease and the outflow decreases.
Thus, the obstacle should be set up the place where it blocks the pedestrians interrupting from the side and does not block the pedestrian walking straight to the exit.
We have discovered this phenomenon since we consider the effect of conflicts and turning by the frictional function and the turning function.

\FIGT{\EPSF{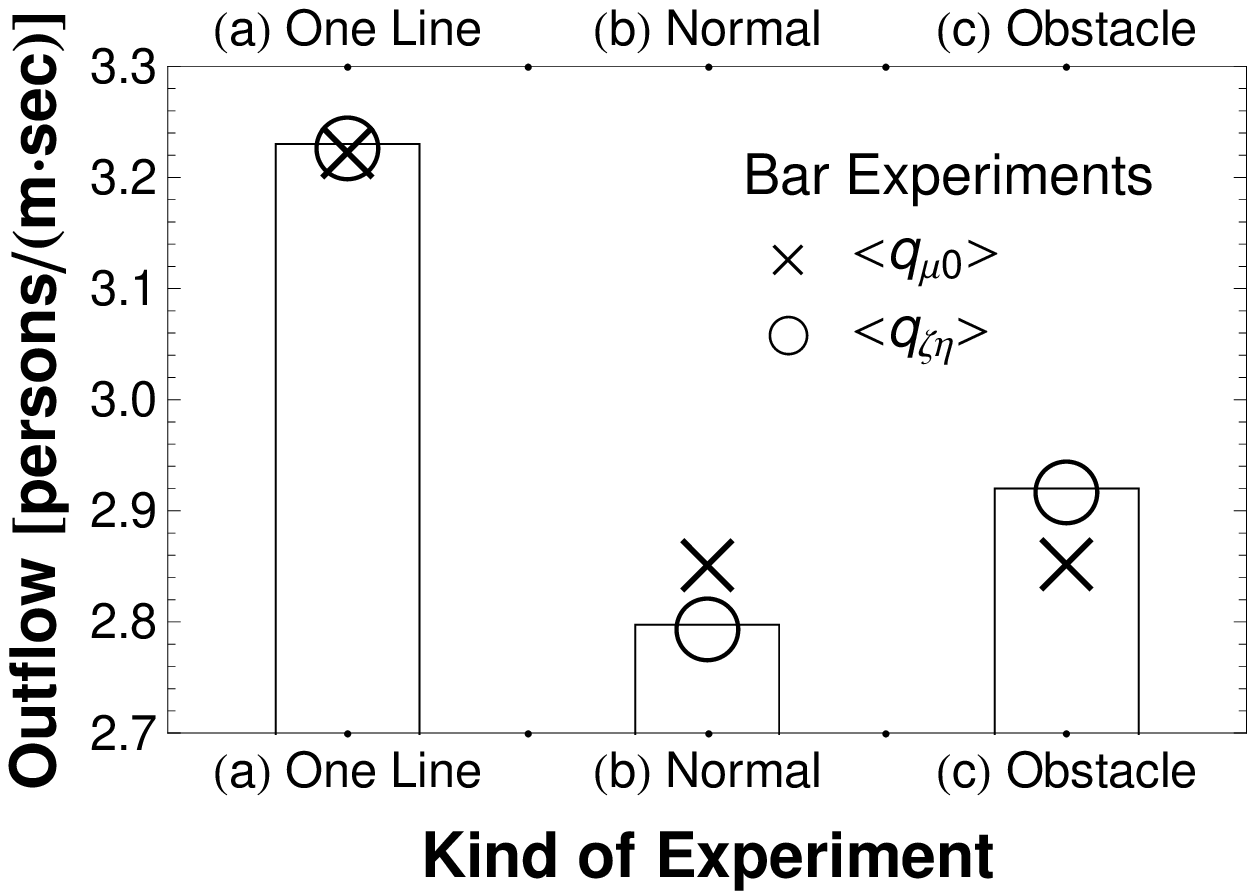}{
The pedestrian outflows from the experiments and the theoretical calculation.
The parameters in Tab. \ref{ana-result1} are used for the theoretical results.
We observe that $\LRA{q_{\zeta \eta}}$ agree with the experimental results well, while $\LRA{q_{\mu 0}}$ does not.
}{height}{5.8}}

\SEC{Conclusion}{CONC}

In this paper we have introduced the \textit{frictional function} and the \textit{turning function} to the floor field model.
The frictional function is the improvement of the friction parameter and changes its value against the number of pedestrians involved in a conflict.
The turning function represents the decrease in walking speed when a pedestrian change their orientation.
Since conflicts and turning are the two major phenomena observed at an exit in evacuation situation, introduction of the two functions enables us to calculate realistic pedestrian outflows through an exit.

We have done the two evacuation experiments.
One is the evacuation in lines in various patterns, and four interesting phenomena have observed.
First, the pedestrian outflow decreases when the number of pedestrians moving to the exit increases since the number and the strength of conflicts increases.
Second, it also decreases when pedestrians need to change their orientation at the exit.
Third, when pedestrians evacuate in two parallel lines, they are intelligent enough to avoid conflicts and smoothly merge into one line.
Forth, we have discovered that approximate four pedestrians are trying to move to the exit at the same time in the normal evacuation when the width of the exit is 50 cm.

By using the frictional function and the turning function, we have succeeded to reproduce the characteristics described in above except the third one and obtained the more realistic figure of the pedestrian outflow through an exit, which corresponds to the result of the experiments very well.
The pedestrian outflow using the friction parameter or excluding the effect of turning does not agree with the experimental results well; therefore, the frictional function and the turning function are necessary for realistic pedestrian outflows.

The other experiment is the evacuation through an exit with an obstacle.
We have clearly shown that the pedestrian outflow increases by putting an obstacle in front of the exit from our experiments.
Our assumption that the pedestrian outflow increases since the obstacle decreases the conflicts at the exit by blocking the pedestrians' movement is verified by theoretical calculation using frictional function.
We have also discovered that the outflow depends on the position of the obstacle.
When the obstacle is shifted from the center of the exit, it blocks the interrupting pedestrians form the side and the outflow increases.
On the contrary, when the obstacle is put at the center, it blocks the pedestrians trying to go through the exit straightly and the outflow decreases.
We have succeeded to obtain this result by introducing the turning function.

While the real pedestrians are intelligent enough to merge smoothly in the evacuation in two lines as we discovered from our experiments, pedestrians in our model are not intelligent enough to do so.
Thus, developing a model which includes such pedestrian intelligence is our future work.
The effect of a size of an obstacle and the width of the exit should be also studied in detail by using both the frictional function and the turning function in the near future.

\section*{Acknowledgments}
We thank the NHK TV program \textit{Science Zero} in Japan for the assistance of the experiment, which is described in Sec. \ref{EXP2}.
This work is financially supported by the Japan Society for the Promotion of Science and the Japan Science and Technology Agency.

\newpage 

\end{document}